\documentclass[aps,pra,reprint]{revtex4-1}

\usepackage{amsmath}
\usepackage{amssymb}
\usepackage{amsfonts}
\usepackage{times}
\usepackage{bm}				
\usepackage{color}			
\usepackage{natbib}
\usepackage[colorlinks]{hyperref}
\usepackage{graphicx} 			

\graphicspath{{Figures/}}

\newcommand{\Grad}{
\vec{\nabla} 
}

\newcommand{\skc}[1]{#1}

\begin{document}

\title{Competing role of interactions in synchronization of exciton-polariton condensates}
\author{Saeed A. Khan, Hakan E. T\"ureci}
\address{Department of Electrical Engineering, Princeton University, Princeton, New Jersey 08544, USA}
\date{\today}

\begin{abstract}
	We present a theoretical study of synchronization dynamics in incoherently pumped exciton-polariton condensates in coupled polariton traps. Our analysis is based on a coupled-mode theory for the generalized Gross-Pitaevskii equation, which employs an expansion in non-Hermitian, pump-dependent modes appropriate for the pumped geometry. We find that polariton-polariton and reservoir-polariton interactions play competing roles and lead to qualitatively different synchronized phases of the coupled polariton modes as pumping power is increased. Crucially, these interactions can also act against each other to hinder synchronization. We map out a phase diagram and discuss the general characteristics of these phases using a generalized Adler equation.
\end{abstract}

\maketitle

Synchronization is a ubiquitous phenomenon in dynamical oscillating systems~\cite{pikovskyBook} that has recently seen renewed interest across diverse setups, from electro/opto-mechanical oscillators~\cite{matheny_phase_2014, zhang_synchronization_2012, holmes_synchronization_2012, bagheri_photonic_2013}, to lasers~\cite{wunsche_synchronization_2005, malik_spectral_2015, bohm_amplitude-phase_2015}, trapped atoms~\cite{hush_spin_2015, lee_quantum_2013, xu_synchronization_2014}, and many-oscillator arrays~\cite{cawthorne_synchronized_1999, cross_synchronization_2004, heinrich_collective_2011}. Recently quantum effects have also been explored~\cite{mari_measures_2013, ludwig_quantum_2013, lee_entanglement_2014, lorch_genuine_2016, lorch_quantum_2017}. In this work, we focus on the manifestation of synchronization in incoherently pumped exciton-polariton systems, where this phenomenon appears as the natural mechanism to explain~\cite{wouters_synchronized_2008} the formation of spatially extended condensates across disorder-generated localized photonic traps, prevalent in early experiments~\cite{baas_synchronized_2008}. It was later pointed out~\cite{eastham_mode_2008} that synchronization can also take place between extended modes that overlap. While such effects are present in photon lasers as well~\cite{lugiato_cooperative_1988, tamm_frequency_1988, tang_synchronization_1996}, what distinguishes polariton condensates are strong nonlinear interactions~\skc{\cite{ge_pattern_2013, galbiati_polariton_2012}}. One may then ask whether these interactions can give rise to {\it qualitatively} different synchronization physics in polariton condensates. Our present work aims to answer this question.

Fundamentally, we study the synchronization of two coupled `oscillators'; however, the unique platform of exciton-polariton condensates under incoherent pumping modifies this picture significantly. In these systems, an uncondensed fraction of polaritons (the ``reservoir'') is deposited by the pump, typically at high energies; interactions amongst them give rise to stimulated scattering towards lower energy states, which can lead to condensation at a threshold pump strength $P_1^{\rm L}$. The condensate mode above this threshold power appears with a self-organized frequency and associated spatial pattern. Typically, above a second threshold $P_2^{\rm L}$ a second mode condenses with generally a different oscillation frequency~\cite{sun_stable_2016, marelic_spatiotemporal_2016}. While this threshold physics is similar to the photon laser, exciton-polaritons distinguish themselves with strong, pump-dependent nonlinear interactions, which come in two varieties. Quasi-particles belonging to the condensate interact, giving rise to purely energetic effects. On the other hand, the deposited reservoir polaritons provide both a repulsive potential and the source of gain that allows condensation in the first place. Crucially, these interactions strongly depend on quasi-particle density, and hence pump power, leading to an effective \textit{pump-dependent} coupling between the oscillators that is different from Huygens' original clock oscillator model of synchronization~\cite{bennett_huygenss_2002, huygens}. The interactions mediate frequency modulation, which we find plays a crucial role in eventually locking the two oscillators to a single frequency at the synchronization threshold. In fact, over the same pumping range, synchronization is no longer observed if interactions are turned off. Strong interactions thus enable the synchronization threshold to be reached even for far detuned modes at low powers; this is different from the laser, where relatively weaker interactions fail to lower the threshold for synchronization below that of other instabilities (e.g. a third mode turning on), except under carefully constructed situations that maximise modal coupling and minimize detuning~\cite{tureci_mode_2005}). 

This simple emergence of synchronization beyond a threshold power is not the full story, however; the different nature and origin of the two distinct nonlinear interactions, a feature quite unique to polariton condensates, reveals even richer dynamics. When one interaction is dominant relative to the other, synchronization is indeed aided by the interaction-mediated frequency modulation. Crucially, the properties of the emergent synchronized state depend strongly on the nature of the dominant interaction. However, we also find that these strong interactions \textit{do not} always help the two oscillators to synchronize. When both interactions have comparable effects - in a quantifiable way that we derive - the situation is reversed: the interactions play competing roles that actually prevent a synchronized state from being reached.

To efficiently capture the nonlinear, pump-dependent interactions and reservoir dynamics present in this system, we employ a temporal coupled mode theory (TCMT) of condensate dynamics introduced in our recent work~\cite{khan_non-hermitian_2016}. This approach is built upon the standard description of polariton condensation via the generalized Gross-Pitaevskii equation (gGPE), but provides a more economical framework for both numerical and analytical studies. In particular, the modal theory enables us to develop an analytical model for the role of interactions based on a multivariable generalized Adler equation, and to numerically study the long-time limit of dynamics for a wide parameter regime. Our key result is a phase diagram which maps out the presence of distinct synchronized phases when either of the two interactions is dominant, and a desynchronized phase when both interactions are actively competing.

The rest of this paper is organized as follows: in Section~\ref{sec:revTCMT}, we briefly review the non-Hermitian coupled-mode theory that is the foundation for the analysis to follow. In Section~\ref{sec:synch}, we introduce the model system for our study of synchronization: two coupled, detuned polariton traps under uniform incoherent pumping. The coupled-mode theory is specialized to this case, and an amplitude-phase-intensity basis is introduced to express the modal equations more efficiently. A generalized Adler equation for the phase dynamics is analyzed in Section~\ref{sec:genAdler}, which provides intuitive insight into the role played by interactions in the existence of a synchronized phase. Finally, in Section~\ref{sec:stability}, the emergence of stable synchronized fixed points as a function of pump power is investigated, and a phase diagram is mapped out. Analytical results are compared both with full numerical simulations of the TCMT, as well as select simulations of the gGPE using a split-step integrator; excellent agreement is found amongst the various methods.

\section{Review of Non-Hermitian Coupled Mode Theory}
\label{sec:revTCMT}

\subsection{Generalized Gross-Pitaevskii Equation}

The dynamics of exciton-polariton condensation under incoherent pumping has been very effectively described via a generalized Gross-Pitaevskii equation (gGPE). This takes the form of a complex dynamical equation for the order parameter $\Psi(\mathbf{r},t)$, coupled to a rate equation for the incoherent reservoir population $n_R$ deposited by the pump (we set $\hbar = 1$)~\cite{wouters_excitations_2007}:
\begin{subequations}
\begin{align}
i\partial_t \Psi &= \left[-\frac{\nabla^2}{2m} + \mathcal{V}(\mathbf{r}) + \left(g_R+i\frac{R}{2}\right)n_R +g|\Psi|^2 -i\frac{\gamma_c}{2} \right]\Psi \label{gGPE} \\
\partial_t n_R &= Pf(\mathbf{r}) -\gamma_Rn_R - Rn_R|\Psi|^2
\label{res}
\end{align}
\end{subequations}
Here, $\mathcal{V}(\mathbf{r})$ is an arbitrary confining potential; for our study of synchronization, $\mathcal{V}(\mathbf{r})$ describes two tunnel-coupled and detuned polariton traps. Such models are relevant to condensation in disordered semiconductor microcavities, as well as recently realized coupled micropillar systems~\cite{abbarchi_macroscopic_2013, baboux_bosonic_2016}. The pump depositing the exciton reservoir has strength $P$ and spatial profile $f(\mathbf{r})$. We consider condensation under uniform incoherent pumping; the confining and pumping geometries are depicted in Fig.~\ref{synchGeo}~(a). 

Scattering from the exciton reservoir provides gain (rate $R$) leading to the formation of a coherent condensate when polariton losses $\gamma_c$ are overcome. The reservoir relaxation rate $\gamma_R$ encapsulates all loss processes other than scattering into the condensate. Most important for the purpose of this work are the two distinct types of interparticle interactions present: between polaritons within the condensate (strength $g$), and between reservoir excitons and condensate polaritons (strength $g_R$). Within the mean-field description employed here, interactions provide repulsive potentials that lead to polaritons experiencing pump-dependent frequency blueshifts. \skc{Note that reservoir-mediated gain and blueshift arise from separate physical processes (scattering and repulsive interactions respectively), and are thus characterized by generally different parameters $R$ and $g_R$.} \skc{Finally}, an additional diffusion term for reservoir excitons may be included in Eq.~(\ref{res}); however, this term is often neglected due to reservoir excitons having much heavier mass relative to that of condensate polaritons. We justify this approximation in Appendix~\ref{app:diff}. 

Unfortunately, the set of coupled nonlinear partial differential equations (PDEs) described by Eqs.~(\ref{gGPE}) are numerically expensive to solve and provide limited analytic insight except in some special cases~\cite{wouters_spatial_2008}. However, from this starting point we are able to arrive at a modal description that proves much more efficient. This approach, originally developed in Ref.~\cite{khan_non-hermitian_2016}, projects Eqs.~(\ref{gGPE}),~(\ref{res}) onto a spatial basis consisting of pump-dependent non-Hermitian modes, which we introduce next. The result of this projection is a set of ordinary differential equations for time-dependent basis expansion coefficients; these equations, expressed in Eqs.~(\ref{an}),~(\ref{Nnm}), comprise our temporal coupled-mode theory (TCMT).

\subsection{Non-Hermitian Pump modes}

To define the spatial basis for the TCMT, we begin by considering \skc{a linearized version of the gGPE, in the absence of polariton-polariton interactions and pump depletion.  We drop nonlinear terms $\propto |\Psi|^2$, and replace the reservoir density by its undepleted value, $n_R = Pf(\mathbf{r})/\gamma_R$; then Eqs~(\ref{gGPE}),~(\ref{res}) reduce to the single linearized dynamical equation:}
\begin{align}
i\partial_t \Psi &= \left[-\frac{\nabla^2}{2m} + \mathcal{V}(\mathbf{r}) + \frac{1}{\gamma_R} \left(g_R + i\frac{R}{2}\right) P f(\mathbf{r}) - i\frac{\gamma_c}{2} \right]\Psi \nonumber \\
&\equiv \mathcal{H}_{\rm L}(P)\Psi
\label{HLP}
\end{align}
\skc{Eq.~(\ref{HLP}) provides an exact description of the physical situation before and \textit{just} upto the formation of a condensate, but is clearly not equivalent to the full gGPE beyond the condensation threshold, where $|\Psi|^2 \neq 0$. For our purposes, however, Eq.~(\ref{HLP}) is used to \textit{define} the linear non-Hermitian generator $\mathcal{H}_{\rm L}(P)$, which incorporates the trapping potential $\mathcal{V}(\mathbf{r})$ and the pump-induced potential. Then, the pump power $P$ is arbitrarily tunable, and serves to parametrize the pump dependence of this linear generator.} The computational modes $\{\varphi_n\}$ we employ for our modal theory are then eigenmodes of this generator, defined as:
\begin{align}
\mathcal{H}_{\rm L}(P)\varphi_n(\mathbf{r};P) = \nu_n(P) \varphi_n(\mathbf{r};P).
\end{align}
\skc{Since $\mathcal{H}_{\rm L}(P)$ is non-Hermitian, its eigenmodes are not orthogonal relative to the standard inner product. However, it can be shown that by introducing a set of dual modes $\{\bar{\varphi}_n\}$ which satisfy the dual non-Hermitian eigenproblem $\mathcal{H}_{\rm L}^*(P) \bar{\varphi}_n = \nu_n^*(P)\bar{\varphi}_n$, a complete basis may be obtained, satisfying the biorthogonality relation~\cite{tureci_self-consistent_2006, ge_steady-state_2010}:
\begin{align}
\int_{\mathcal{P}} d\mathbf{r}~\bar{\varphi}_n^*(\mathbf{r};P)\varphi_m(\mathbf{r};P) = \delta_{nm},
\label{biorth}
\end{align}
where $\mathcal{P}$ is the minimal region beyond which the pump has vanishing strength. It is easy to see that $\bar{\varphi}_n^* = \varphi_n$, so that Eq.~(\ref{biorth}) reduces to a self-orthogonality relation}:
\begin{align}
\int_{\mathcal{P}} d\mathbf{r}~\varphi_n(\mathbf{r};P)\varphi_m(\mathbf{r};P) = \delta_{nm}
\label{orth}
\end{align}

\skc{The linear dynamics described by Eq.~(\ref{HLP}) are} encoded in the pump-dependent, complex eigenvalues $\nu_n(P) = \omega_n(P) + i\gamma_n(P)$ \skc{of each eigenmode}. The real part $\omega_n(P)$ represents the modal frequency, determined by the confining potential $\mathcal{V}(\mathbf{r})$ and the blueshift due to reservoir excitons ($\propto g_R$). On the other hand, the imaginary parts $\gamma_n(P)$ of the eigenvalues characterize non-Hermitian pumping and dissipation: they describe the net gain experienced by the $n$th mode at a given pump power. For low enough pump powers, all modal eigenvalues have negative imaginary parts, indicative of a below-threshold regime - all modes experience net loss. As the pump power increases, the modal eigenvalues flow in the complex plane as the gain from the pump increases. For a specific pump strength $P = P_n^{\rm L}$, which we refer to as the linear threshold power, the $n$th eigenmode acquires a vanishing imaginary part; for $P > P_n^{\rm L}$, this mode experiences gain [See Fig.~\ref{synchGeo}~(c)]. A threshold pump power can be associated with each mode; it is determined by how effective the specific mode is at utilizing gain from the pump~\cite{khan_non-hermitian_2016}, in the presence of the confining and pump-induced potential. Modes that are spatially configured to utilize the pump more efficiently have lower threshold pump values, and vice versa. 

Due to their explicit pump dependence and associated threshold physics, we refer to the non-Hermitian eigenmodes of the linear generator as \textit{pump modes} from this point on. Furthermore, the linear threshold power value $P_n^{\rm L}$ associated with each pump mode $\{\varphi_n\}$ hints at a natural ordering principle for these non-Hermitian modes. This allows us to a define a suitable truncation scheme for an expansion in a basis of these pump modes, which we discuss next.

\subsection{Coupled-Mode Equations}

Having defined our computational basis, we are now equipped to tackle the full nonlinear problem posed by Eqs.~(\ref{gGPE}),~(\ref{res}). We proceed to expand the condensate order parameter $\Psi(\mathbf{r},t)$ in the set of pump-dependent non-Hermitian pump modes with \textit{time-dependent} coefficients:
\begin{align}
\Psi(\mathbf{r},t) = \sum_{n=1}^N a_n(t) \varphi_n(\mathbf{r};P)
\label{exp}
\end{align}
where the sum may extend over an arbitrary number $N$ of pump modes. The choice of modes to include in this expansion may seem unclear. Note that a truncation scheme that retains only pump modes with the lowest frequencies would not be appropriate in all cases: the non-Hermitian dynamics can lead to condensation in a mode that is \textit{not} the lowest frequency mode for a given pumping potential. Instead, a truncation based on the linear threshold powers of the pump modes is more appropriate. We choose a basis of size $N$ to comprise of the modes with the $N$ \textit{lowest} linear threshold power values. In this way, the modes that best optimize gain from the pump are retained in the basis expansion~\cite{khan_non-hermitian_2016}.

With the expansion placed on firm footing, we may now project the fully nonlinear Eqs.~(\ref{gGPE}),~(\ref{res}) onto this basis, and integrate out the spatial dependence using Eq.~(\ref{orth}). We leave the details of this procedure for Appendix~\ref{app:TCMT}; the resulting dynamical equations corresponding to Eq.~(\ref{gGPE}) and Eq.~(\ref{res}) respectively are given by:
\begin{subequations}
\begin{align}
&i \frac{d}{dt} a_n = \nu_n(P)a_n  +\!\left(g_R + i\frac{R}{2}\right) \sum_{m} N_{nm}a_m \nonumber \\
&~~~~~~~~~~~~~~~+ g \sum_{mkq} A_{nmkq}a_ma_ka_q^* \label{an} \\
&\frac{d}{dt}N_{nm} = -\gamma_R N_{nm} \nonumber \\
&-\! \sum_{kq}\! \left\{ \! \left[\frac{R}{\gamma_R}P B_{nmkq} + A_{nmkq} \left( i\nu_k - i\nu^*_q \right)  \right] + \left(\frac{d}{dt}\right)\! \right\}a_ka_q^*
\label{Nnm}
\end{align}
\end{subequations}
In the above, we have introduced dynamical variables describing the evolving reservoir density, referred to as the reservoir matrix elements $N_{nm}(t)$. These are given by:
\begin{align}
N_{nm} = \int_{\mathcal{P}} d\mathbf{r}~\varphi_n\left[ n_R - \frac{P}{\gamma_R}f(\mathbf{r})\right]\varphi_m,
\label{resElems}
\end{align}
and hence describe the change in reservoir density relative to its unsaturated value. Also, we define the spatial overlap matrix elements:
\begin{align}
A_{nmkq} = \! \int_{\mathcal{P}} d\mathbf{r}~\varphi_n\varphi_m\varphi_k\varphi_q^*~,~B_{nmkq} = \! \int_{\mathcal{P}} d\mathbf{r}~f(\mathbf{r})\varphi_n\varphi_m\varphi_k\varphi_q^*
\label{Anmrs}
\end{align}
Eqs.~(\ref{an}),~(\ref{Nnm}) constitute the TCMT with nonlinear interactions and pump depletion. In Eq.~(\ref{an}), we clearly see the eigenvalues $\nu_n(P) = \omega_n(P) + i \gamma_n(P)$ controlling linear dynamics. All remaining terms incorporate the nonlinear effects neglected in the linear theory; the term $\propto g$ describes polariton-polariton interactions, with the mode overlap integrals $A_{nmkq}$ describing self-interaction and cross-mode interaction contributions. The final term with reservoir matrix elements $N_{nm}$ describes the effect of reservoir depletion; note that these reservoir matrix elements are themselves dynamically determined by Eq.~(\ref{Nnm}).

The TCMT provides an efficient spectral description of the nonlinear, non-Hermitian dynamics of polariton condensation, for arbitrary pumping and trapping potentials. In Ref.~\cite{khan_non-hermitian_2016}, TCMT simulations were compared with direct split-step integration of the gGPE, and the results found to agree very favourably in a variety of settings. In the next section, we employ this TCMT for an analytical and numerical study of synchronization in coupled polariton traps.

\section{Condensation in Coupled Polariton Traps}
\label{sec:synch}

\subsection{System and Parameters}
\label{subsec:params}

\begin{figure}[t]
\begin{center}
\includegraphics[scale=0.27]{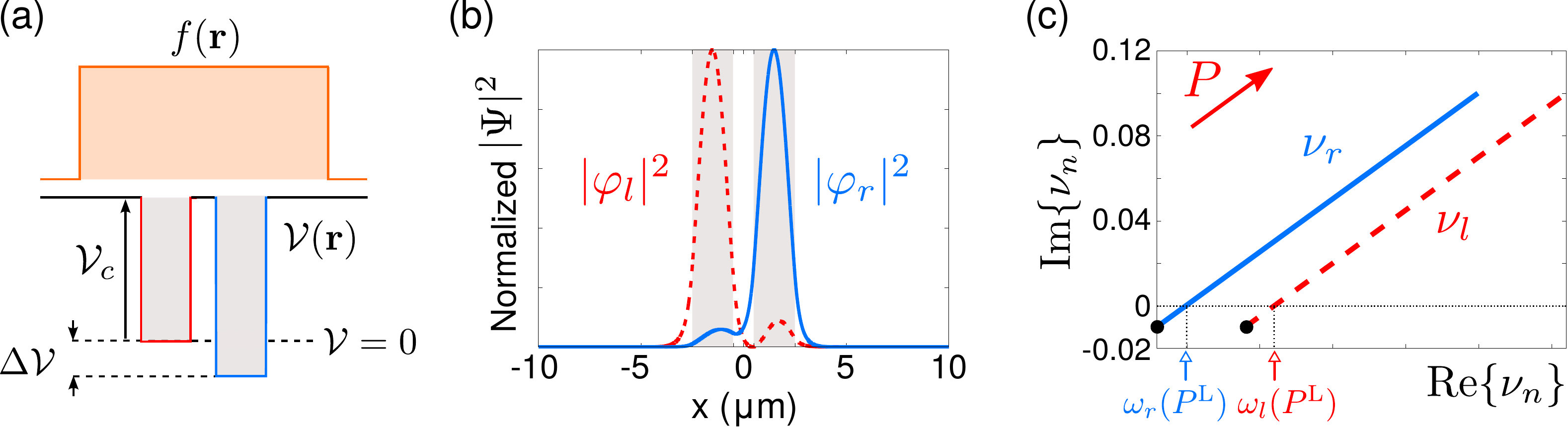}
\caption{(a) Trapping potential $\mathcal{V}(\mathbf{r})$ and pump profile $f(\mathbf{r})$ for the study of synchronization. (b) Density profile of the two non-Hermitian pump modes used as the basis for the two-mode TCMT. (c) Evolution of the associated complex-valued eigenvalues $\nu_n (P)$ with pump power in the complex $\nu$ plane.}
\label{synchGeo}
\end{center}
\end{figure}

For the study of synchronization, we consider a confining potential $\mathcal{V}(\mathbf{r})$ that describes two coupled polariton traps, and a uniform incoherent pump across both traps [See Fig.~\ref{synchGeo}~(a)]. Generally, there is a (low) pumping range wherein it is accurate to truncate the TCMT to only the two most preferred pump modes, namely the modes with lowest linear threshold powers. We consider specifically a detuned trap regime here, where these two modes are predominantly confined to the left or right trap; hence we denote the modes by $\varphi_{l}$ and $\varphi_r$ respectively, as shown in Fig.~\ref{synchGeo}~(b). We emphasize again that these modes are \textit{not} simply eigenmodes of the individual traps; rather they account for tunneling between the traps, the unsaturated pump, and dissipation through cavity loss. \skc{Under pumping with a wide, homogeneous pump spot, it is the static trapping potential that predominantly determines the modes $\varphi_l$, $\varphi_r$; the non-Hermitian pump-induced potential does not modify the modes directly in this case, but our analysis does not make any explicit simplifications using this fact. However the non-Hermitian potential controls the evolution of modal eigenvalues with pump power:} the homogeneous pump spot ensures this evolution is similar for both eigenvalues [see Fig.~\ref{synchGeo}~(c)]. Furthermore, both modes have an approximately equal linear threshold power $P^{\rm L}$; this is evident in how the eigenvalues cross the real line concurrently as pump power is increased. In the absence of pump-induced repulsion, the modal frequencies are $\omega_{l0}$ and $\omega_{r0}$, determined by the trapping potential alone; we choose $\omega_{l0} > \omega_{r0}$, and define the bare pump mode detuning $\Delta\omega \equiv \omega_{l0}-\omega_{r0}$; for the potential landscape $\mathcal{V}(\mathbf{r})$ chosen here, $\Delta\omega_0 = 0.12~{\rm meV}$, a typical value consistent with modal detunings in a variety of polariton trapping structures~\cite{baas_synchronized_2008, baboux_bosonic_2016}. For concreteness, we set $R = 0.1~\mu{\rm m^2~meV}$, $m^{-1} = 0.59~\mu{\rm m^2~meV}$, $\gamma_c = 1~{\rm meV}$, and $\gamma_R = 10~{\rm meV}$. 

Note that we have chosen to operate in the regime of fast reservoir relaxation, $\gamma_R \gg \gamma_c$, which is relevant for condensation in disordered semiconductor quantum wells. More importantly, in the opposite case where $\gamma_R \ll \gamma_c$, we find a tendency towards strongly multimode behaviour, in agreement with recent numerical studies in this \skc{non-adiabatic regime}~\cite{bobrovska_stability_2014, bobrovska_adiabatic_2015, khan_non-hermitian_2016}. The complicated dynamics can in fact be captured quite effectively by the TCMT, provided we include more than two modes in our basis~[See Appendix~\ref{app:dynRes}]. Therefore, to study dynamics within a two-mode approximation, we restrict ourselves to the fast reservoir relaxation regime.

\subsection{Amplitude-Phase-Intensity basis}

\skc{The condensate wavefunction in the two-mode TCMT is given by Eq.~(\ref{exp}) for $n = l,r$, with the time-dependence entirely included in the mode coefficients $\{a_n(t)\}$. For the study of synchronization, it proves useful to access the amplitude and phase dynamics directly; to this end we write the coefficients in a phasor representation, $a_n(t) = \bar{a}_n(t) e^{-i\phi_n(t)},~n = l,r$, and then cast Eqs.~(\ref{an})~,(\ref{Nnm}) in terms of the dynamical variables $\{\phi,z,\rho\}$: }
\begin{align}
\phi = \phi_l - \phi_r~,~z = \frac{\bar{a}_l^2-\bar{a}_r^2}{\bar{a}_l^2+\bar{a}_r^2}~,~\rho = \bar{a}_l^2 + \bar{a}_r^2
\label{tcmtVars}
\end{align}
where $\phi$ is the relative phase, $z$ the normalized modal intensity imbalance, and $\rho$ the total intensity. Note that both $\phi \in [-\pi,\pi]$, and $z \in [-1,1]$ are \skc{bounded} variables, while $\rho$ generally increases monotonically with the pump power $P$; these observations prove very useful in our analysis later. Also, in the regime of fast reservoir relaxation, the reservoir matrix elements $N_{nm}$ can be solved for in terms of the coefficients $\{a_n\}$, so that the $\{\phi,z,\rho\}$ variables are sufficient for a full description of the reservoir-condensate system.

A large body of earlier work on synchronization~\cite{acebron_kuramoto_2005} concerns itself with weakly coupled Hermitian systems: the total intensity $\rho$ is a conserved quantity and oscillator amplitudes are deemed to be approximately stationary so that $\dot{z} \approx 0$, thereby allowing a phase-only description of the dynamics. \skc{In the desynchronized regime, the modes typically evolve at two distinct frequencies, namely the blueshifted trap frequencies, but modified by nonlinear interactions. The phases $\phi_{l,r}(t)$ then exhibit drift-like evolution. On the other hand, in the synchronized regime the two modes begin oscillating at the \textit{same} unique frequency $\Omega_0$. When this occurs, both phases $\phi_{l,r}(t)$ exhibit a linear drift in time with the same drift constant, given by the synchronized frequency, $\Omega_0 t$, plus (in general) a constant offset. Crucially, then, it follows that the \textit{relative} phase $\phi = \phi_l-\phi_r$ becomes \textit{stationary} in time, $\dot{\phi} = 0$. This requirement is an analytic signature of frequency synchronization in such phase-only systems.}

For the non-Hermitian, pump-dependent system under consideration here, we find it crucial to keep track of $z$ and $\rho$ dynamics in addition to the relative phase evolution. Oscillations in $z$ and $\rho$ are pump-dependent and may be large~[See Appendix~\ref{app:sysDyn}], so that the aforementioned approximations are not always valid. More importantly, while the relative phase may become stationary momentarily ($\dot{\phi} = 0$), \skc{we find that} this state can only persist if $\dot{z} = \dot{\rho} = 0$ concurrently, and if the state is stable to fluctuations in \textit{any} of these dynamical quantities. \skc{The requirement of such stable fixed points in $\{\phi,z,\rho\}$ space is then the analytic signature of synchronization; comparisons between analytic and numerical results in Section~\ref{subsec:phase} confirm that the modal frequency detuning does in fact vanish at such fixed points. 
}

\skc{Although the present system is not described by a phase-only model, the phase dynamics still remain quite informative:} since $\dot{\phi} = 0$ is still a necessary (though not sufficient) condition for synchronization to occur, it can place strong constraints on the synchronized phase, which we will explore next.

\section{Generalized Adler Equation}
\label{sec:genAdler}

One of the earliest models of synchronization involves the standard single-variable Adler equation~\cite{adler_study_1946}:
\begin{align}
\dot{\phi} = \Omega - F \sin(\phi)
\label{adler}
\end{align}
The Adler equation describes the dynamics of the relative phase $\phi$, of two coupled oscillators for example, in the presence of a detuning term $\Omega$ that causes $\phi$ to drift linearly in time, and a coupling term $F\sin(\phi)$ that encourages its pinning to a constant value. Clearly, the synchronized $\dot{\phi}=0$ solution is possible only if $-F < \Omega < F$, namely when the detuning term is small enough compared to the coupling. This model very successfully describes a broad range of synchronization and phase locking dynamics, from injection locking of oscillators to an external drive~\cite{razavi_study_2004, bhansali_gen-adler:_2009}, to the synchronization of coupled oscillators within a phase-only picture~\cite{mirzaei_mutual_2014}.

In the present case, a somewhat more complicated equation for the relative phase $\phi$ may also be obtained by transforming Eqs.~(\ref{an}),~(\ref{Nnm}) to $\{\phi,z,\rho\}$ space as defined in Eq.~(\ref{tcmtVars}); we refer to it as the generalized Adler equation:
\begin{align}
\dot{\phi} = \Omega(z,\rho) - \rho F(\phi,z).
\label{genAdlerFull}
\end{align}
In analogy with the standard Adler equation, we refer to the $\phi$-independent term $\Omega(z,\rho)$ as the detuning term (generally \textit{not} just the same as the bare pump mode detuning $\Delta\omega_0$), and $\rho F(\phi,z)$ as the coupling term, which is periodic in $\phi$. The $\rho$ dependence of both terms reflects the pump-dependent nature of the nonlinear modal interaction. Note that $\Omega$ in Eq.~(\ref{adler}) represents the bare frequency detuning of the \textit{uncoupled} oscillators, and not the actual emergent detuning in the presence of coupling. Similarly, the detuning term $\Omega(z,\rho)$ is \textit{not} the emergent detuning between the two coupled polariton modes; the latter is determined via simulation of the modal equations~[See Section~\ref{subsec:phase}].

The condition for $\dot{\phi}=0$ imposed by the generalized Adler equation becomes:
\begin{align}
{\rm min_{\phi}}\{\rho F(\phi,z)\} < \Omega(z,\rho) <  {\rm max_{\phi}}\{\rho F(\phi,z)\}
\label{genAdlerCond}
\end{align}
where ${\rm max_{\phi}/min_{\phi}}\{f\}$ are the maximum/minimum values respectively taken by $f(\phi,z,\rho)$ as $\phi$ varies in $[-\pi,\pi]$, for a given $(z,\rho)$. Eq.~\ref{genAdlerCond}) implies that synchronization is again possible only when the detuning is small enough compared to the coupling term; now, however, these terms are no longer constants like in Eq.~(\ref{adler}), but are rather configuration dependent. To understand the implications of this constraint, we study the detuning and coupling terms separately, beginning with the former. 

\subsection{Nonlinear Detuning modification}
\label{subsec:detuning}

\begin{figure}
\begin{center}
\includegraphics[scale=0.22]{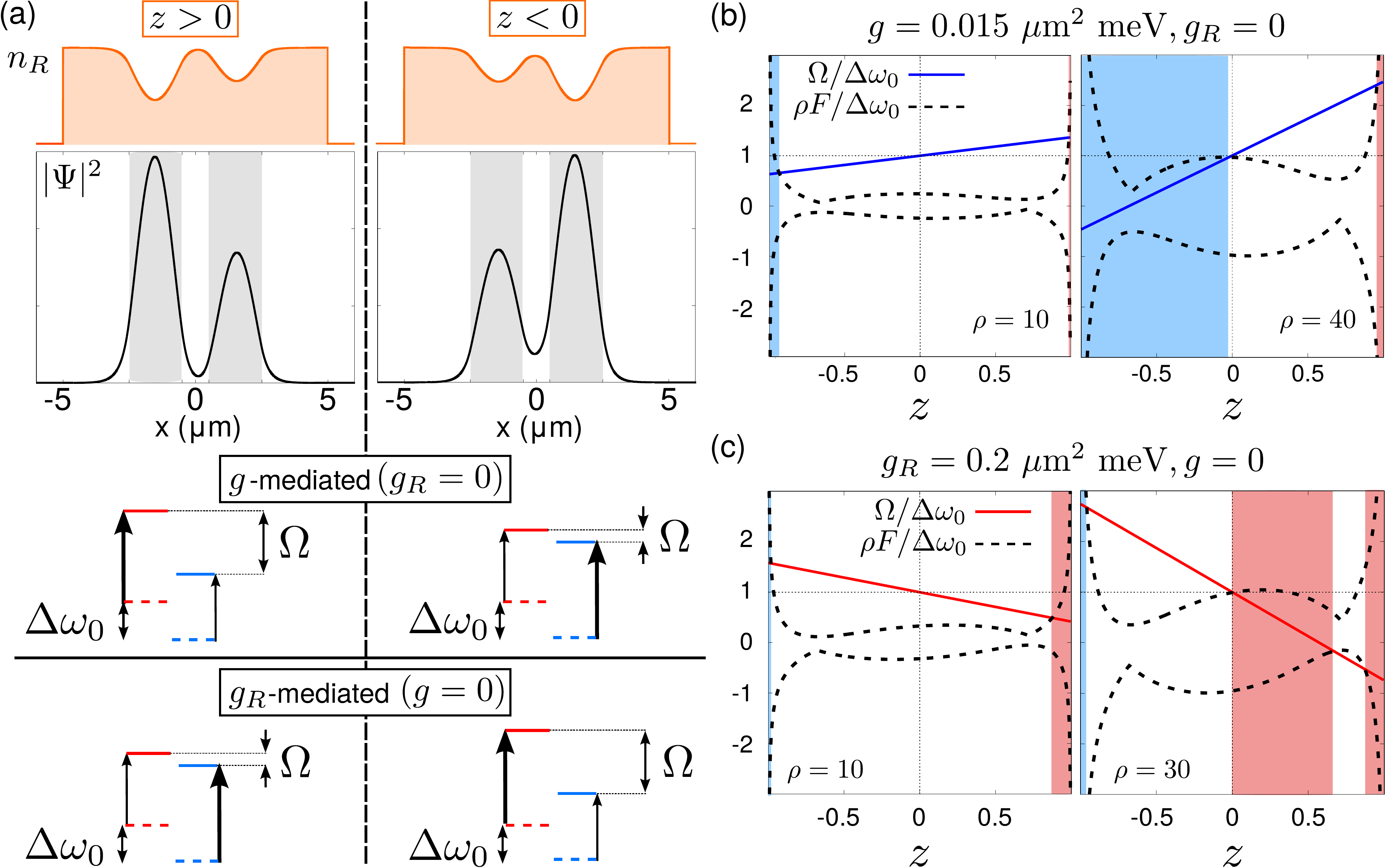}
\caption{(a) Schematic variation of detuning term $\Omega$ with $z$ in the $g$-mediated regime (middle) and $g_R$ mediated regime (bottom). In the $g$-mediated regime, the frequency blueshift comes from polaritons within the condensate. Hence the effective modal detuning term $\Omega$ is lowered when the low frequency mode has higher occupation ($z<0$) and experiences a stronger blueshift. In the $g_R$-mediated regime, the blueshift is now due to reservoir excitons, and $\Omega$ is instead reduced when the low frequency mode has lower occupation ($z>0$) and sees a less depleted exciton reservoir. Right panel: Evolution of detuning (solid blue/red) and coupling terms (dashed black), scaled by $\Delta\omega_0$, with $\rho$ in the (b) $g$-mediated regime ($g_R=0$), and (c) the $g_R$-mediated regime ($g=0$). Synchronization is possible only in the shaded regions. These regions prefer $z < 0$ in the $g$-mediated regime and $z>0$ in the $g_R$-mediated regime, and grow with $\rho$.}
\label{detuning}
\end{center}
\end{figure}

The detuning term $\Omega(z,\rho)$ includes the bare modal frequency difference and its modification due to nonlinear interactions. Here, we will see that this detuning term provides intuitive insight into the effect of nonlinear interactions on synchronization dynamics. The full expression for the detuning term is given by Eq.~(\ref{app:Omega}) in Appendix~\ref{app:fullDyn}; in the main text, we find it more instructive to present the form $\Omega(z,\rho)$ takes for specific regimes. In the absence of any repulsive interactions, $g,g_R=0$, the detuning term is a constant equal to the bare pump mode detuning,
\begin{align}
\Omega = \Delta\omega_0~~~~~~~~~~~~~~~[g= g_R = 0]
\end{align}
independent of the pump power. The situation becomes more interesting once either type of repulsive interaction is active. If only interactions of polaritons within the condensate are considered - we define this as the $g$-mediated regime ($g_R = 0$) - the detuning term becomes:
\begin{align}
\Omega =~~&\left(\omega_{l0} + g A_{llll}\bar{a}_l^2 + 2g A_{llrr}\bar{a}_r^2\right) \nonumber \\
- &\left(\omega_{r0} + g A_{rrrr}\bar{a}_r^2 + 2g A_{rrll}\bar{a}_l^2\right)~~~~~~~[g_R = 0]
\label{detG}
\end{align}
We have momentarily reverted to amplitude variables $\bar{a}_n$ since each term is intuitively clearest here. The first bracketed term represents the blueshift of the the left trap mode due to repulsion from polaritons occupying that mode ($\propto A_{llll}\bar{a}_l^2$), and from polaritons occupying the right trap mode ($\propto A_{llrr}\bar{a}_r^2$); the latter arises since the modes have nonzero spatial overlap. The second bracketed term is the corresponding blueshift of the right trap mode. Here $\frac{A_{llrr}}{A_{llll}}\approx \frac{A_{rrll}}{A_{rrrr}} \simeq 0.1$, so that the `direct' blueshift is more important, and causes the mode with higher occupation to be more strongly blueshifted due to $g$. A reduction in the detuning term - which aids synchronization, as per Eq.~(\ref{genAdlerCond}) - thus requires the low frequency mode ($r$) to experience a stronger blueshift, which is possible when this mode has higher occupation ($\bar{a}_r^2 > \bar{a}_l^2$), namely $z < 0$, as depicted in Fig.~\ref{detuning}~(a).

In the $g_R$-mediated regime where only interactions between reservoir excitons and condensate polaritons are active ($g = 0$), we have instead:
\begin{align}
\Omega = &\left( \omega_l(P)  +  g_R \frac{\rho\gamma_c}{\gamma_R} N_{ll}(z)  \right) \nonumber \\
-&\left( \omega_r(P) +   g_R \frac{\rho\gamma_c}{\gamma_R} N_{rr}(z) \right)~~~~~~~[g = 0]
\label{detGR}
\end{align}
The exciton repulsion $\propto g_R$ now linearly blueshifts the mode frequencies, lending them the $P$ dependence shown in Fig.~\ref{synchGeo}~(c); the homogeneous pump spot ensures an equal blueshift that leaves the pump mode detuning unchanged, $\omega_l(P)-\omega_r(P)=\Delta\omega_0$. However, condensate formation depletes the exciton reservoir, since the latter serves as the source of gain. In particular, the higher the occupation of a mode, the more depleted is the exciton reservoir it sees. This reservoir depletion reduces the reservoir-mediated blueshift experienced by each mode; in the detuning term, this reduction is captured by the $N_{nn}$ elements, which measure the reservoir depletion seen by mode $n$, and are explicitly negative [c.f. Eq.~(\ref{resElems})]. To reduce the detuning term in this case, the low frequency mode ($r$) again needs to be more strongly blueshifted; however, due to the different blueshift mechanism, the lower frequency mode ($r$) must now have a \textit{lower} occupation ($\bar{a}_r^2 < \bar{a}_l^2$), namely $z>0$, since it then sees a less depleted exciton reservoir and can experience a stronger blueshift. This is exactly the opposite to the $g$-mediated case [see Fig.~\ref{detuning}~(a)]. Hence, we see a configuration-dependence to the reduction of the detuning term in the generalized Adler equation, with the $g$- and $g_R$-mediated regimes preferring opposite configurations.

Finally, note the additional prefactor of $\gamma_c/\gamma_R = 0.1$ relative to the $g$-mediated case, which weakens the $g_R$ mediated blueshift. This can be explained as follows: for larger values of $\gamma_R$, the pump threshold for condensation increases. The stronger pumping allows for a stronger scattering $\propto R|\Psi|^2$ from the reservoir to the condensate before reservoir depletion becomes important, which in turn means a higher condensate occupation is possible (for fixed polariton loss rate $\gamma_c$). The saturated reservoir density, however, is unchanged, since the stronger scattering uses up the additional pump power. Therefore, reservoir-dependent terms are reduced by a factor of $\gamma_c/\gamma_R$ relative to terms that depend on the condensate density alone.

\subsection{Pump-dependent coupling}
\label{subsec:coupling}

Moving on to the coupling term $\rho F(\phi,z)$, our choice of variables immediately indicates its growth with pump power via the explicit scaling by $\rho$; the precise dependence is clarified later. More explicitly, one finds:
\begin{align}
F(\phi,z) = \frac{1}{\sqrt{1-z^2}} \left[ gF_g(\phi,z) - g_R N_{rl} \left( \frac{\gamma_c}{\gamma_R} \right)F_{g_R}(\phi,z)\right]\!\!,
\label{coupling}
\end{align}
a nonlinear function of $z$ with multi-harmonic $\phi$-dependence. Note that the coupling term may also be divided into a $g$-mediated term $\propto F_g$ and a $g_R$-mediated term $\propto F_{g_R}$; the latter is again weaker in this regime by the factor $\gamma_c/\gamma_R$. The explicit forms of these functions may be found in Appendix~\ref{app:fullDyn}. Most importantly, we find the coupling term is approximately unchanged under $z\to -z$, which is very different from the behaviour of the detuning term.

Both detuning and coupling terms combine to determine the possibility of a synchronized phase via Eq.~(\ref{genAdlerCond}). This is best explored graphically; we first consider the $g$-mediated regime, where $g_R = 0$. In Fig.~\ref{detuning}~(b), we plot ${\rm max_{\phi}/min_{\phi}}\{\rho F(\phi,z)\}$ (dashed black) and $\Omega(z,\rho)$ (solid blue) for fixed $\rho$, over the entire range of $z \in [-1,1]$. The value of $\rho$ is proportional to the pump power: the left panel is for $\rho = 10$, while the right panel shows $\rho=40$, corresponding to an increased pump power. Eq.~(\ref{genAdlerCond}) for $\dot{\phi} =0$ is satisfied in the shaded regions; here the detuning term lies within the range of the coupling term as determined by the dashed curves. Hence, in the \textit{unshaded} regions, synchronization is impossible; clearly, the $z<0$ region is preferred for synchronization here, a result stemming from the reduced value of $\Omega$ for this configuration. With increasing pump power, the shaded region area grows, as the increasing coupling strength and detuning modification make synchronization easier. The analogous plot in the $g_R$-mediated regime ($g= 0$) is shown in Fig.~\ref{detuning}~(c), for $\rho = 10$ and $\rho = 30$. Here, the situation is effectively reversed: the different mechanism for frequency modification means that the detuning term (solid red) is reduced for $z>0$ instead, whereas the coupling term (dashed black) is mostly unchanged. As a result, synchronization is preferred here for the $z>0$ configuration.

\section{Stability Analysis and Phase Diagram}
\label{sec:stability}

\subsection{Stability Maps}
\label{subsec:maps}

The intuitive description of the previous section places constraints on the $\dot{\phi}=0$ state, but does not guarantee synchronization; a complete analysis requires studying the full system of equations given by:
\begin{subequations}
\begin{align}
\dot{\phi} &= 0 =~\Delta\omega_0 -\rho G_{\phi}(\phi,z) \label{phiFull} \\
\dot{z} &= 0 =~ \left(\gamma_l-\gamma_r\right)(1-z^2) +  \rho G_z(\phi,z)  \label{zFull} \\
\dot{\rho} &= 0 =~\rho\left[ (\gamma_l + \gamma_r) + (\gamma_l-\gamma_r)z -\rho G_{\rho}(\phi,z) \right] \label{rhoFull}
\end{align}
\end{subequations}
Here we have rewritten Eq.~\ref{genAdlerFull} to isolate the bare pump mode detuning, $\Delta\omega_0$. $G_{\phi,z,\rho}$ are functions of system parameters and the variables $(\phi,z)$ only. The explicit forms of these functions are provided in Appendix~\ref{app:fullDyn}; we will find that the expressions above can already yield useful insight. While Eqs.~(\ref{phiFull})-(\ref{rhoFull}) cannot be analytically solved for the fixed points, progress can be made if both modal gains are taken to be equal, $\gamma_l(P) \approx \gamma_r(P)$. Recall that the $\gamma_n(P)$ are pre-determined by solving the non-Hermitian problem for the pump modes; we find the modal gains are indeed numerically equal in the present case. Note that this is also the most interesting scenario, since neither mode is preferred over the other by the pump.

We consider fixing all system parameters other than the pump power. The $\dot{\rho}$ equation can then be solved to obtain the parametric dependence (on $\phi$ and $z$) of the fixed points of $\rho$, 
\begin{align}
\rho_{\rm FP} = \frac{\gamma_l(P) + \gamma_r(P)}{G_{\rho}(\phi,z)}
\end{align} 

The dominant power dependence here comes from the evolution of $\gamma_n(P)$, which is shown in Fig.~\ref{synchGeo}~(c). The function $G_{\rho}(\phi,z)$ in the denominator is independent of $P$, and hence for a given $(\phi,z)$ pair, the steady state value of $\rho$ evolves {\it linearly} with pump power. In what follows, it is then justified to use $\rho$ as a surrogate variable for the pump power $P$. The $\dot{z}$ equation also simplifies to $\dot{z} = 0 = \rho~G_z(\phi,z)$. This implies that the values of $(\phi,z)$ for which $\dot{z}$ vanishes are unchanged with increasing $\rho$, and hence pump power. 

In addition to the existence of fixed points in $\{\phi,z,\rho\}$ space, a persistent synchronized phase demands that such fixed points be dynamically stable. We find that a standard stability analysis of Eqs.~(\ref{phiFull})-(\ref{rhoFull}) based on the eigenvalues $\lambda$ of the associated Jacobian matrix $\mathbf{J}$ also benefits from the simplified $\rho$ dependence. At the fixed points, we have:
\begin{align}
\mathbf{J}(\phi,z,\rho)\Big|_{\rm FP} = \rho 
\left. \begin{pmatrix}
-\partial_{\phi}G_{\phi} & -\partial_z G_{\phi} & -\frac{\Delta\omega_0}{\rho^2} \\
\partial_{\phi}G_z & \partial_z G_z & 0 \\
-\rho\partial_{\phi}G_{\rho} & -\rho\partial_zG_{\rho} & -G_{\rho} 
\end{pmatrix}\right|_{\rm FP}
\label{jac}
\end{align}
The matrix element $\propto \Delta\omega_0/\rho^2$ yields terms in the characteristic equation that are suppressed by $1/\rho$ relative to remaining contributions at the same order. For large enough $\rho$, this term may be dropped; then, the resulting Jacobian matrix has a characteristic equation $\chi(\phi,z,\lambda/\rho) = 0$, with a crucial implication: the $\rho$ dependence of the characteristic equation serves only to scale its roots, the eigenvalues of $\mathbf{J}$. This result is explicitly derived in Appendix~\ref{app:jac}. Hence, the \textit{sign} of any eigenvalue does not change as $\rho$, and therefore pump power, is changed. While this conclusion holds only approximately, and that too at the fixed points of Eqs.~(\ref{phiFull})-(\ref{rhoFull}) at any pump power, we find that in practice the signs of eigenvalues of $\mathbf{J}$ are quite robust to changes in $\rho$, for values of $\rho$ close to $\rho_{\rm FP}$ at that pump power. Thus, for any fixed set of system parameters, stable regions where all eigenvalues of $\mathbf{J}$ have negative real parts, and unstable regions where at least one eigenvalue has positive real part, are approximately {\it unchanged} with pump power. Because $\phi$-$z$ space is bounded in $[-\pi,\pi]\times[-1,1]$, the emergence and movement of fixed points with increasing pump power can be tracked on finite, fixed \textit{stability maps} of stable and unstable regions to characterize the global behavior of the system (even though the entire space is not explored for any given initial condition).

\begin{figure}[t]
\begin{center}
\includegraphics[scale=0.265]{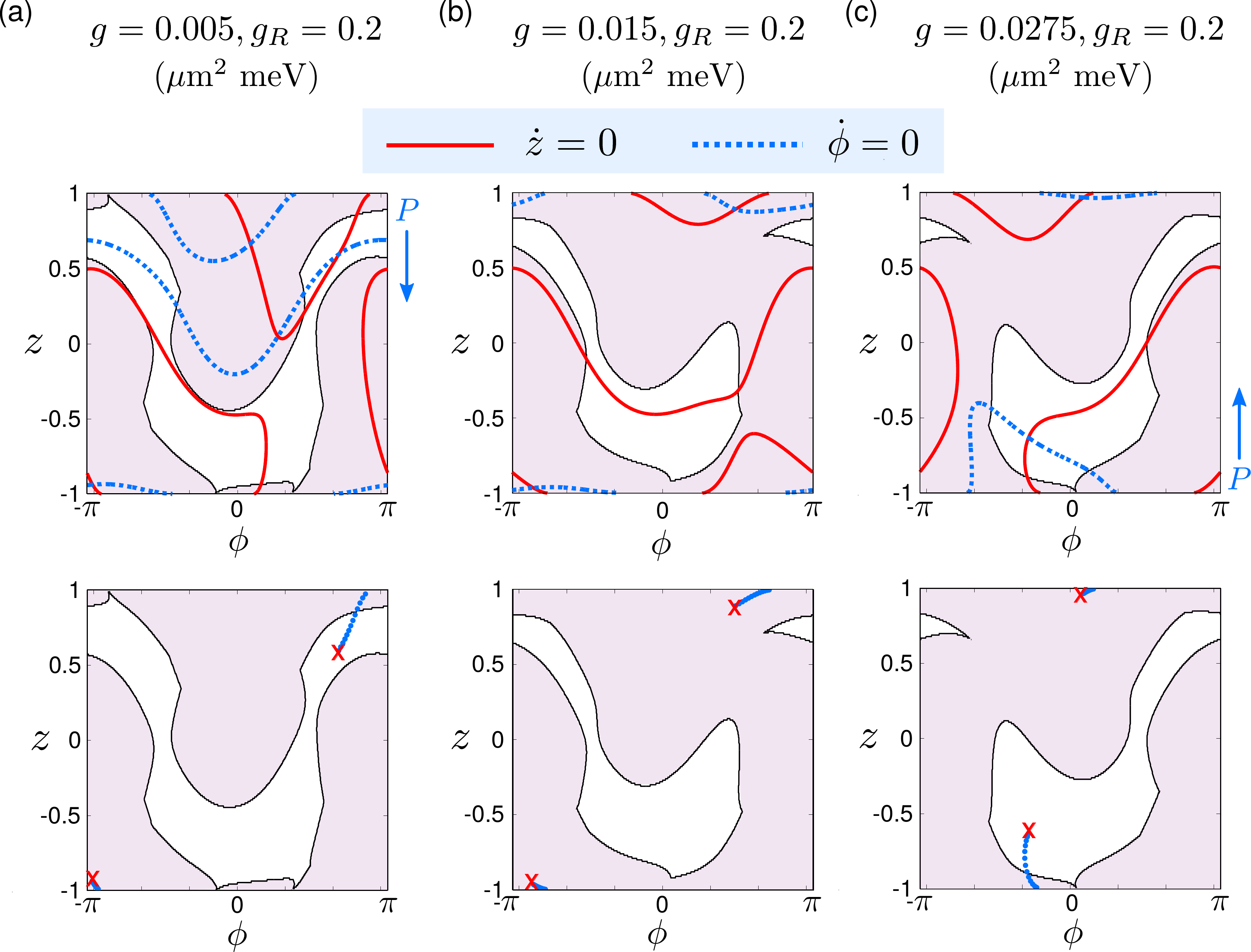}
\caption{Top row: Stable (plain) and unstable (shaded) regions, with $\dot{z}=~0$ contours (solid red) and $\dot{\phi} = 0$ contours (dashed blue), for (a) $g_R$-mediated synchronization, (c) $g$-mediated synchronization, and (b) desynchronized regime due to $g$-$g_R$ competition. Corresponding interaction strengths are indicated above each plot. Arrows in (a), (c) indicate movement direction of $\dot{\phi}=0$ contours; the depicted contours were plotted for $P = 1.15P^{\rm L}$ [Supplementary animation of this panel can be found online]. Bottom row: Corresponding evolution of fixed points on stability maps as pump power is increased, $P\in[1,1.15]P^{\rm L}$. Red cross marks final position of a fixed point.}
\label{stabilityAnalysis}
\end{center}
\end{figure}

An example of a stability map is shown for the $g_R$-mediated regime in Fig.~\ref{stabilityAnalysis}~(a), with plain regions being stable and shaded regions unstable, and the $\dot{z}=0$ contours shown in solid red; all of these features are unchanged with pump power. Only the $\dot{\phi}=0$ contours (dashed blue) change with $P$; note that such a contour only exists at points where condition~(\ref{genAdlerCond}) is satisfied. Hence we can now see how the generalized Adler equation plays a defining role in the emergence of a synchronized phase with increasing pump power. In the $g_R$-mediated regime, it is clear from Fig.~\ref{detuning}~(c) that the $\dot{\phi} = 0$ contours prefer the $z>0$ region. The trajectories of fixed points with $P$ are shown in the lower panel of Fig.~\ref{stabilityAnalysis}~(a). The fixed point in the $z>0$ region flows with increasing $P$ and enters the stable region; as this crossing occurs, a synchronized phase of the system becomes stable. Note that in the $z<0$ region, the $\dot{z}=0$ contour exists in a stable region; however, a stable fixed point cannot emerge until the $\dot{\phi}=0$ contour spreads in this region, which is clearly restricted based on our analysis of the generalized Adler equation [See Fig.~\ref{detuning}~(c)]. In contrast, Fig.~\ref{stabilityAnalysis}~(c) shows a stability map in the $g$-mediated regime. Here, the $\dot{\phi}=0$ contours prefer instead the $z<0$ region, again clear from Fig.~\ref{detuning}~(b). As such, the fixed point that flows from an unstable to a stable region now has $z<0$.

\begin{figure*}[t]
\begin{center}
\includegraphics[scale=0.28]{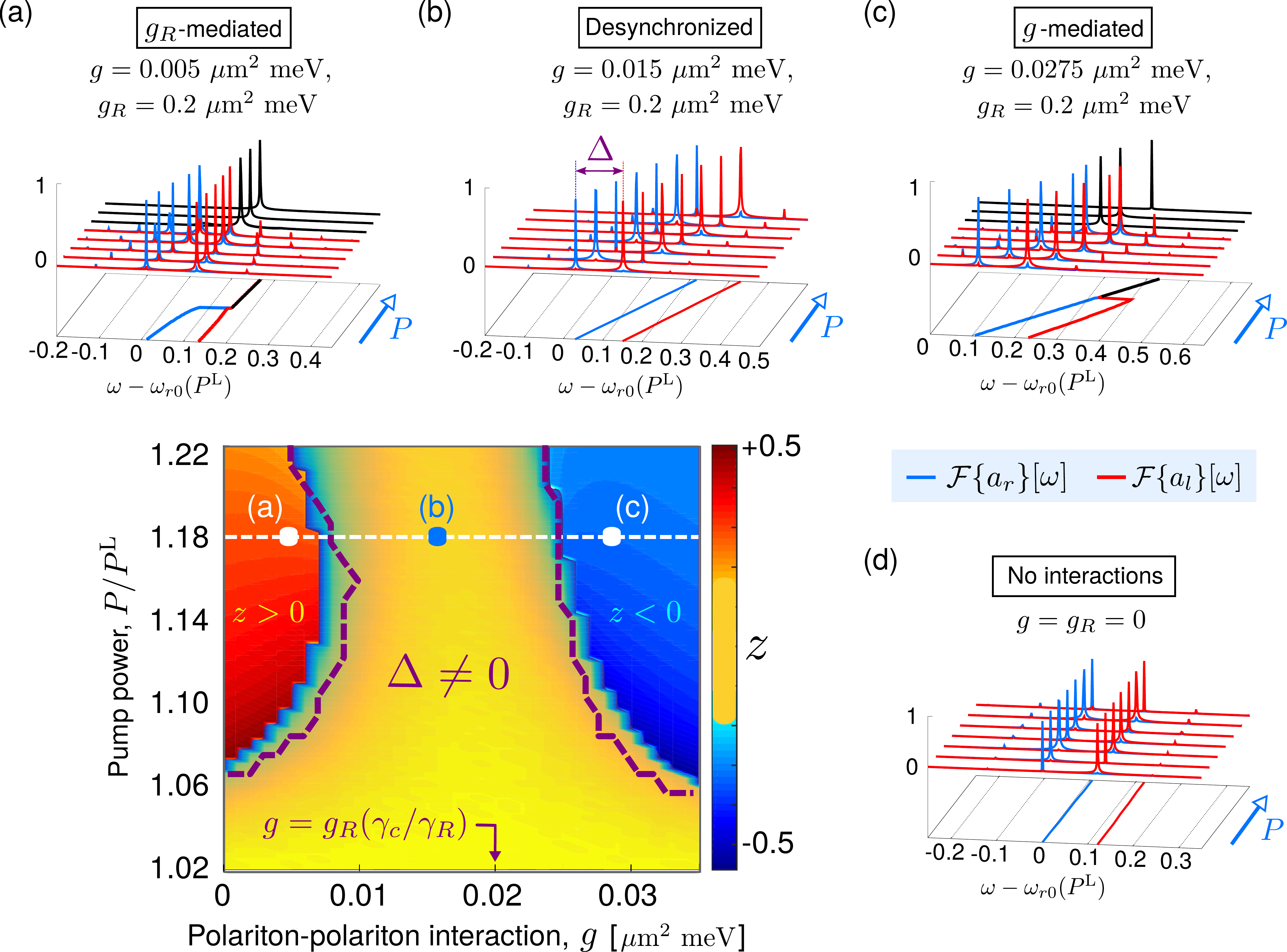}
\caption{Phase diagram in $P$-$g$ space for $g_R =~0.2~\mu{\rm m^2}~{\rm meV}$, showing a $g_R$-mediated synchronized regime with $z > 0$ \skc{(shaded red)} and a $g$-mediated regime with $z<0$ \skc{(shaded blue)}, separated by a desynchronized regime \skc{(shaded yellow)}. Also shown is the evolution with pump power of typical frequency spectra $\mathcal{F}\{a_n\}$ in (a) the $g_R$-mediated regime, (b) the desynchronized regime, and (c) the $g$-mediated regime. Projections below the spectra track the evolution with pump power of the largest frequency peak of each mode.}
\label{phaseDiagComb}
\end{center}
\end{figure*}

Most interestingly, an intermediate regime exists where the two interactions compete; the different scaling of detuning and coupling terms discussed earlier implies that for this regime, $g\sim g_R (\gamma_c/\gamma_R)$. A typical stability map here is shown in Fig.~\ref{stabilityAnalysis}~(b). For the same range of pump strengths as Fig.~\ref{stabilityAnalysis}~(a),~(c), the $\dot{\phi}=0$ contours barely move; this is due simply to the competing effects of $g$- and $g_R$-mediated frequency modification. As a result, while there are stable regions where $\dot{z}=0$ contours also exist, these are not accessible to the almost static $\dot{\phi}=0$ contours. Hence, no stable fixed point emerges and a desynchronized phase persists [Supplementary animation showing the movement of $\dot{\phi} =0$ contours in Fig.~\ref{stabilityAnalysis} can be found online].

\subsection{Phase Diagram}
\label{subsec:phase}

\begin{figure*}[t]
\begin{center}
\includegraphics[scale=0.35]{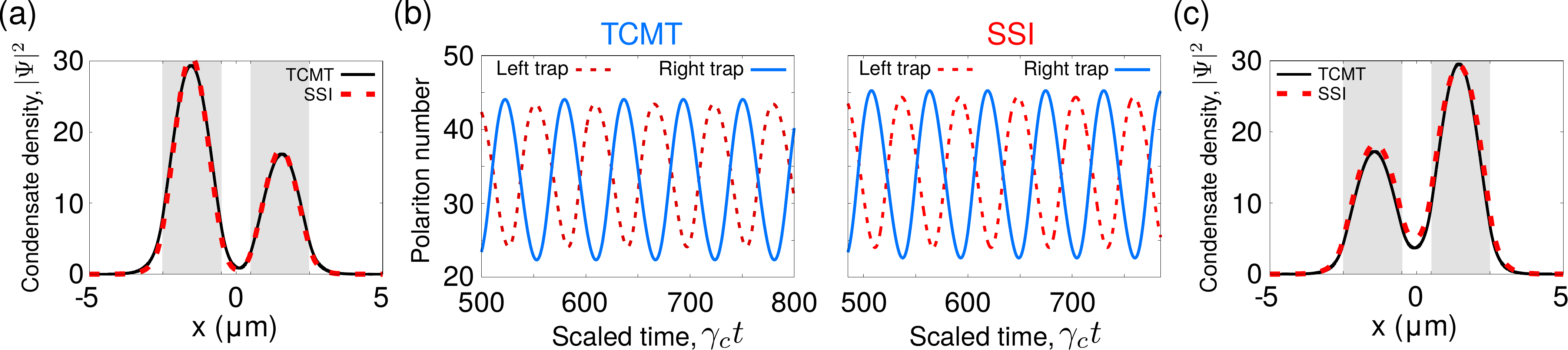}
\caption{(a), (c) Steady state condensate density $|\Psi|^2$ (solid black) at correspondingly labelled positions in the two synchronized regions of the phase diagram in Fig.~\ref{phaseDiagComb}, together with the SSI results (dashed red). (b) Oscillating polariton number in each trap in the correspondingly labelled desynchronized region of the phase diagram, computed using the TCMT and SSI.}
\label{SSIComparison}
\end{center}
\end{figure*}

The predictions of the previous sections manifest strikingly in a phase diagram obtained via full numerical integration of the TCMT. We simulate the dynamical mode coefficients $a_n(t)$ to long times until a steady state is reached, and then compute the modal detuning $\Delta$ from the Fourier transform of mode coefficients, $\mathcal{F}\{a_n\}$. This procedure is repeated for varying pump strengths $P$ and increasing values of the polariton-polariton interaction strength $g$, starting from $g=0$; importantly, the reservoir-polariton interaction strength is kept \textit{fixed} at $g_R=0.2~\mu{\rm m^2~meV}$. A plot of the computed detuning $\Delta$ in $P$-$g$ space constitutes the phase diagram shown in Fig.~\ref{phaseDiagComb}. \skc{In the yellow shaded regions, a desynchronized solution persists ($\Delta \neq 0$). Outside this region, the modal detuning vanishes and the two modes synchronize. Three distinct dynamical regimes can be identified, which we will now describe. For weak enough values of $g\ll g_R(\gamma_c/\gamma_R)$, a synchronized state emerges beyond a threshold pump power. In this region, the typical pump dependence} of the Fourier spectra $\mathcal{F}\{a_n\}$ of mode coefficients is shown in Fig.~\ref{phaseDiagComb}~(a). The curves projected below the spectra depict the evolution of the mode frequencies as a function of pump power. Clearly, the modes experience a frequency blueshift due to interactions as the pump is increased (dotted lines indicate constant frequency). This continues until the threshold power \skc{is reached,} beyond which a single frequency synchronized state emerges \skc{(indicated in solid black)}. \skc{In the synchronized case, $\dot{\phi} = \dot{z} = \dot{\rho} = 0$ in the steady state; we thus superimpose the steady state population imbalance in this synchronized regime on Fig.~\ref{phaseDiagComb}. For weak $g$, the polariton configuration appears with $z > 0$ (shaded red), in what we label as the regime of $g_R$-mediated synchronization.} 

\skc{With} stronger $g \sim g_R\left(\gamma_c/\gamma_R\right)$, the predicted competition between the two types of frequency modulation effects does in fact arise, and the synchronized state disappears. Typical frequency spectra $\mathcal{F}\{a_n\}$ in this regime, shown in Fig.~\ref{phaseDiagComb}~(b), still indicate a frequency blueshift due to interactions, \skc{but the modes remain detuned.} \skc{In this interacting regime, then, dynamics are reminiscent of ac Josephson oscillations between the coupled condensates [See Section~\ref{subsec:compSSI}].} For values of $g$ that are stronger still, a synchronized phase emerges once more, but now with $z < 0$ \skc{(shaded blue)}. In this $g$-mediated regime where interactions are strongest, the frequency blueshift is most pronounced, as is clear from the Fourier spectra in Fig.~\ref{phaseDiagComb}~(c). Again, the blueshifted modes lock to a single frequency beyond a synchronization power threshold. 

The threshold for synchronization in both the $g_R$-mediated and $g$-mediated regimes is predicted very well by our analysis of fixed points with \skc{$\dot{\phi} = \dot{z} = \dot{\rho} = 0$} moving on a fixed stability map. The stability map is computed at fixed $P=1.15P^{\rm L}$, making use of our previous result that stable and unstable regions are approximately unchanged with pump power. The phase boundary computed via this analysis is shown in dashed purple in Fig.~\ref{phaseDiagComb}. \skc{The slight discrepancy near the phase boundary may be explained as follows. For pump powers above the dashed purple line, the synchronized solution becomes stable; however, in a narrow range of pump powers, the initial condition determines whether the system settles into the synchronized phase or remains desynchronized. We note that this bistability of synchronized and desynchronized solutions is similar to results found in another two-mode configuration by Borgh \textit{et. al.}~\cite{borgh_spatial_2010}.}

Interestingly, for the pump range studied here, a synchronized state does not emerge if both interactions are turned off. For this case where $g=g_R=0$, the Fourier spectra of mode coefficients as a function of pump power are shown in Fig.~\ref{phaseDiagComb}~(d). The modes experience no blueshift, as expected, and the modal detuning is unchanged with pump power. Therefore, while the presence of either a dominant polariton-polariton interaction or reservoir-polariton interaction is necessary for synchronization, the competition of \textit{both} interactions actually hinders synchronization. 

\skc{Finally, one may note that the value of $g_R$ used in Fig.~\ref{phaseDiagComb} is strong relative to $g$; this is chosen to make clear the role played by reservoir-polariton interactions in synchronization. Phase diagrams for weaker $g_R$ values are included in Appendix~\ref{app:resDyn}, Fig.~\ref{fig:suppDet}. The $g_R$-mediated synchronization region generally shrinks and can even disappear for weaker values of the reservoir-polariton interaction strength. However, the key element - the competing nature of the interactions - still remains: if for $g_R = 0$, synchronization occurs at a given pump strength for a fixed value of $g$, then nonzero $g_R$ pushes up this pump strength for $g$-mediated synchronization, and for strong enough $g_R$ may even lead to $g_R$-mediated synchronization instead.} 

\skc{
\subsection{Comparisons with SSI}
\label{subsec:compSSI}
}

\skc{Since} the TCMT is derived from the generalized Gross-Pitaevksii equation, a direct comparison between TCMT simulations \skc{of the previous section} and a symplectic split-step integration (SSI) of the gGPE [c.f. Eqs.~(\ref{gGPE}),~(\ref{res})] can be made. \skc{In the one-dimensional geometry under consideration, we find that the two-mode TCMT is between one to two orders of magnitude faster than the SSI at a given pump power, for an equivalent spatial resolution and final integration time. The efficiency is primarily due to the TCMT's avoidance of spatial integration at every time step via a modal expansion. This fact also saves the TCMT's computation times from an unfavourable scaling with spatial dimension. SSI computation times, on the other hand, scale exponentially with dimension $d$, as $N_g^d \log (N_g)$ for a spatial grid with (uniform) density $N_g$~\cite{bader_solving_2013}, which should render the TCMT even more favourable in higher dimensions (not considered here).} 

\skc{Thus, instead of computing the full phase diagram using the SSI, we provide} select comparisons of TCMT and SSI results; in particular this is done for distinct points along the horizontal dashed white line in Fig.~\ref{phaseDiagComb}, which stretches across the three distinct dynamical regimes. In the synchronized regimes, at positions labelled (a) and (c) on the phase diagram, \skc{the condensate density is stationary in the steady state as mentioned earlier. We thus compare $|\Psi|^2$, shown in Figs.~\ref{SSIComparison}~(a),~(c) respectively}; TCMT results are in solid black, and SSI results in dashed red. \skc{The synchronized configurations in (a), (c) are distinguished from self-trapping in coupled condensates~\cite{abbarchi_macroscopic_2013} within the $\phi$-$z$ plane: while frequency synchronization leads to a constant relative phase (and population imbalance) as discussed earlier, the self-trapped regime has a relative phase that drifts linearly in time, and population imbalance that is weakly oscillating~\cite{albiez_direct_2005}.}

In the desynchronized regime, \skc{the mode amplitudes contain multiple frequencies unlike the synchronized case [See Fig.~\ref{phaseDiagComb}~(b)]. The condensate density is then no longer stationary in time. Here a plot of the polariton occupations in each trap, in Fig.~\ref{SSIComparison}~(b), indicates ac Josephson-like oscillations between the detuned traps, with a running phase~\cite{wouters_synchronized_2008, borgh_spatial_2010, abbarchi_macroscopic_2013, voronova_detuning-controlled_2015, rahmani_polaritonic_2016}. The TCMT and SSI exhibit excellent agreement in all three cases}, which helps validate the \skc{preceding} results of the two-mode TCMT. 

\skc{Lastly, the dynamics of the reservoir across the synchronized and desynchronized regimes may also be computed using both methods; these results are included in Appendix~\ref{app:resDyn} for completeness, and are also found to agree well.}   \\

\section{Conclusions \skc{and Outlook} }

In this paper we have investigated the synchronization of two detuned, coupled polariton modes as a function of pump power, with a particular focus on the role played by the different nonlinear interactions unique to pumped polaritonic systems. Our analysis is based on a temporal coupled-mode theory that allows both analytical insight and efficient numerical simulations. We find that the polariton configuration in the emergent synchronized phase is strongly dependent on the relative influence of polariton-polariton and reservoir-polariton interactions. Most interestingly, the two types of interactions can have competing effects that prevent the emergence of a synchronized phase altogether. This conclusion is verified against direct simulation of the generalized Gross-Pitaevksii equation and very good agreement is found.

\skc{
A natural extension of this work is to larger coupled systems, such as polariton lattices~\cite{milicevic_edge_2015}. In particular, in recently realized flat band condensation~\cite{baboux_bosonic_2016}, polariton states are highly localized due to disorder, while still possessing a finite frequency dispersion. The original experiment was concerned with threshold physics under weak pumping, and thus the polariton-polariton nonlinearity played a negligible role. However under stronger pumping, the relative strengths of reservoir-polariton and polariton-polariton interactions could be crucial in determining whether or not a spatially extended, synchronized (single-frequency) condensate could be formed, and what spatial configuration such a condensate may take. For studies in modern polariton lattice structures, the work here can be extended to regimes where $\gamma_R \lesssim \gamma_c$; the regimes directly studied in this paper should still be relevant in disorder-generated polariton trap geometries~\cite{baas_synchronized_2008, wouters_excitations_2007}. More generally, our results uncover an additional role of reservoir-mediated interactions in condensate dynamics, adding to other effects studied recently including the dynamical reservoir regime, multimode dynamics, and instabilities~\cite{bobrovska_stability_2014, bobrovska_adiabatic_2015, khan_non-hermitian_2016}.

}

\section*{Acknowledgements}

We thank F. Marquardt for helpful discussions. This work was supported by the US Department of Energy, Office of Basic Energy Sciences, Division of Materials Sciences and Engineering under Award No. DE-SC0016011.

\begin{center}
\rule{5cm}{1pt}
\end{center}

\appendix

\section{Additional details of TCMT derivation}
\label{app:TCMT}

To project Eqs.~(\ref{gGPE}),~(\ref{res}) onto the non-Hermitian TCMT basis, it is useful to first rewrite Eq.~(\ref{gGPE}) under a displacement transformation of the reservoir density:
\begin{align}
n_R(\mathbf{r},t) = \widetilde{n}_R(\mathbf{r},t) + \frac{P}{\gamma_R}f(\mathbf{r})
\label{disp}
\end{align}
Here, $\widetilde{n}_R(\mathbf{r},t)$ is the nonlinear, time-dependent part of the reservoir density that describes the depletion of the linear part. Under this transformation, and using the definition of $\mathcal{H}_{\rm L}(P)$ in Eq.~(\ref{HLP}), the gGPE can be rewritten as:
\begin{align}
i\partial_t \Psi = \left[ \mathcal{H}_{\rm L}(P) +s\gamma_R\widetilde{n}_R(\mathbf{r},t) + g|\Psi|^2  \right]\Psi
\end{align}
This equation can then simply be projected onto the basis modes $\{\varphi_n\}$ using the expansion in Eq.~(\ref{exp}) and the orthogonality relation of Eq.~(\ref{orth}). The resulting modal equations for the coefficients $a_n(t)$ then become Eq.~(\ref{an}) of the main text. 

The dynamical equation for the full reservoir density $n_R$, Eq.~(\ref{res}), requires some more work. In particular, we employ the continuity equation for this non-Hermitian system, which has the form:
\begin{align}
\vec{\nabla}\cdot\vec{j} + \partial_t |\Psi|^2 = Rn_R|\Psi|^2 -\gamma_c|\Psi|^2
\end{align}
where $\vec{j}$ is the polariton current,
\begin{align}
\vec{j} = \frac{i}{2m} \left( \Psi \Grad{\Psi}^* - c.c.\right)
\end{align}
The left hand side of the continuity equation is simply the continuity equation for a conserved system, while the right hand side describes modifications due to the non-Hermitian nature of this driven-dissipative system. These include gain from the reservoir $\propto R$, and losses $\propto \gamma_c$. Here, we employ the continuity equation to rewrite the reservoir-condensate coupling term in Eq.~(\ref{res}) in terms of condensate-only terms. The rewritten reservoir dynamics equation following this substitution and the displacement transformation of Eq.~(\ref{disp}) takes the form:
\begin{align}
\partial_t \widetilde{n}_R = - \gamma_R\widetilde{n}_R - \partial_t |\Psi|^2 - \gamma_c|\Psi|^2 - \vec{\nabla} \cdot \vec{j}
\end{align}
The above equation is now in a convenient form to be projected onto the non-Hermitian basis. Again employing the basis expansion in Eq.~(\ref{exp}), then multiplying through by $\varphi_n\varphi_m$ and integrating over $\mathcal{P}$, we arrive at Eq.~(\ref{Nnm}) for the dynamical reservoir matrix elements.

\section{Full equations for dynamical variables}
\label{app:fullDyn}

We now present the full forms of the dynamical equations for variables $\{\phi,z,\rho\}$, obtained from Eqs.~(\ref{an}),~(\ref{Nnm}) via the transformation defined in the main text, Eq.~(\ref{tcmtVars}). In what follows, the overlap matrix elements $A_{nmrs},B_{nmrs}$ are as defined in Eq.~(\ref{Anmrs}), while the reservoir matrix elements are scaled to extract out the explicit dependence on $\rho$, $N_{nm} \to \rho\frac{\gamma_c}{\gamma_R}N_{nm}$. 

We begin with the equation for the relative phase $\phi$,
\begin{align}
\dot{\phi} = \Delta\omega_0 - \rho G_{\phi}(\phi,z) 
\label{app:phi}
\end{align}
The bare detuning $\Delta\omega_0 = \omega_l(P)-\omega_r(P)$, as presented in the main text. The function $G_{\phi}(\phi,z)$ can be written as:
\begin{align}
G_{\phi}(\phi,z) = -\frac{1}{2} \left[ \left(\frac{\gamma_c}{\gamma_R}\right)N_{\phi}(\phi,z) + gK_{\phi}(\phi,z) \right]
\end{align}
Here, $N_{\phi}(\phi,z)$ contains the reservoir matrix elements:
\begin{align}
N_{\phi}(\phi,z) =~&2g_R \left( N_{ll}\!-\!N_{rr} \right)  \nonumber \\ -&\frac{1}{\sqrt{1-z^2}} N_{rl} \left( 2R \sin \phi\! +\! 4 z g_R \cos \phi \right)
\end{align}
The function $K_{\phi}(\phi,z)$ describes polariton-polariton repulsion terms:
\begin{align}
K_{\phi}(\phi,z) = &\frac{\cos\phi}{\sqrt{1-z^2}} \Big[ (1-z^2)\left(A_{lllr}\! +\! 2A_{llrl}\! -\! A_{rrrl}\! -\! 2A_{rlrr}\right) \nonumber \\
&~~~~~~~~~~~- (1+z)^2A_{rlll}\!+\!(1-z)^2A_{lrrr}  \Big] \nonumber \\ 
&\!\!+\cos 2\phi  \left[ (1-z)A_{lrrl}\!-\! (1+z)A_{rllr}\right] \nonumber \\
&\!\!+(1+z)\left(A_{llll}\!-\!2A_{rrll}\right)\! -\! (1-z)\left(A_{rrrr}\! -\! 2A_{llrr}\right) 
\end{align}
To compare with the form of the generalized Adler equation of Section~\ref{sec:genAdler}, the full expression for the detuning term $\Omega(z,\rho)$ contains simply the $\phi$-\textit{independent} terms from Eq.~(\ref{app:phi}):
\begin{align}
\Omega(z,\rho) =&~\Delta\omega_0 + g_R \rho \left(\frac{\gamma_c}{\gamma_R}\right) \Big[ N_{ll}(z) - N_{rr}(z) \Big] \nonumber \\
+\frac{1}{2} g \rho \Big[ &(1+z)(A_{llll} - 2A_{rrll}) - (1-z)(A_{rrrr} - 2A_{llrr}) \Big] 
\label{app:Omega}
\end{align}
Note that $N_{nn}(z)$ is the $\phi$-\textit{independent} part of the reservoir matrix element $N_{nn}$. These parts will be made clear in due course. The coupling term of Eq.~(\ref{coupling}) contains all the $\phi$-\textit{dependent} terms from Eq.~(\ref{app:phi}):
\begin{align}
F(\phi,z) = \frac{1}{\sqrt{1-z^2}} \left[gF_g(\phi,z) - g_R N_{rl}(\phi,z) \frac{\gamma_c}{\gamma_R} F_{g_R}(\phi,z) \right] 
\end{align}
Here, $N_{rl}(\phi,z)$ is the $\phi$-\textit{dependent} part of the matrix element $N_{rl}$. The functions $F_g$ and $F_{g_R}$ are defined as:
\begin{align}
F_g(\phi,z) &= \cos\phi \Big[ (1-z^2)\left(A_{lllr} + 2A_{llrl} - A_{rrrl} - 2A_{rlrr}\right) \nonumber \\
&~~~~~~~~~~- (1+z)^2A_{rlll} +(1-z)^2A_{lrrr}  \Big] \nonumber \\ 
&~+\cos 2\phi \sqrt{1-z^2} \left[ (1-z)A_{lrrl}- (1+z)A_{rllr}\right] \nonumber \\
F_{g_R}(\phi,z) &= \left( \frac{R}{g_R} \sin \phi + 2 z \cos \phi \right)
\end{align}

Next, we move on to the dynamical equation for the modal intensity imbalance, $z$, defined in the main text as:
\begin{align}
\dot{z} = (\gamma_l-\gamma_r)(1-z^2) - \rho G_z(\phi,z)
\end{align}
Here, the function $G_z(\phi,z)$ has the form:
\begin{align}
G_z(\phi,z) = - \frac{1}{2} \left[ \left(\frac{\gamma_c}{\gamma_R}\right)N_z(\phi,z) + g K_z \phi,z) \right]
\end{align}
As before, $N_z(\phi,z)$ contains the reservoir matrix elements:
\begin{align}
N_z(\phi,z) =~&(1-z^2) R \left( N_{ll}-N_{rr} \right)  \nonumber \\
-&\sqrt{1-z^2} N_{rl} \left( 2R z \cos \phi -4 g_R \sin \phi \right)
\end{align}
while the function $K_z(\phi,z)$ describes polariton-polariton repulsion terms:
\begin{align}
K_z(\phi,z) =~& \sin 2\phi \Big[A_{rllr}\left( 1+z-z^2-z^3\right) \nonumber \\
&~~~~~~~+ A_{lrrl}\left(1-z-z^2+z^3 \right) \Big] - \nonumber \\
\sqrt{1-z^2} &\sin\phi \Big[ (1-z^2) \left(A_{rrrl} \!-\! A_{lllr}\! -\! 2 A_{rlrr}\! +\! 2A_{llrl} \right) \nonumber \\
&~~~~~~-(1+z)^2 A_{rlll}   - (1-z)^2A_{lrrr} \Big]
\end{align}

Lastly, we move on to the dynamical equation for $\dot{\rho}$:
\begin{align}
\dot{\rho} &= \rho\left[ (\gamma_l + \gamma_r) + (\gamma_l-\gamma_r)z - \rho G_{\rho}(\phi,z) \right]
\end{align}
Now, $G_{\rho}(\phi,z)$ is:
\begin{align}
G_{\rho}(\phi,z) = - \frac{1}{2}\left[ \left(\frac{\gamma_c}{\gamma_R}\right)N_{\rho}(\phi,z) + gK_{\rho}(\phi,z)\right]
\end{align}
where
\begin{align}
N_{\rho}(\phi,z) = R\left[(1+z) N_{ll}\!+\!(1-z)N_{rr}\! +\! 2 \sqrt{1-z^2}N_{rl} \cos \phi \right] 
\end{align}
and 
\begin{align}
K_{\rho}(\phi,z) =~&\sin 2\phi (1-z^2)\left(A_{lrrl} -A_{rllr}\right)~+ \nonumber \\
\sqrt{1-z^2} &\sin\phi\Big[ (1-z)(A_{rrrl}\! +\! A_{lrrr})\! -\! (1+z)(A_{rlll}\!+\!A_{lllr}) \nonumber \\
&~~~- 2(1-z)A_{rlrr}+2(1+z)A_{llrl} \Big]
\end{align}
In all of the above, note that the reservoir matrix elements $N_{nm}$ themselves are dynamical quantities that can be expressed in terms of the variables $(\phi,z)$. In the regime where $\gamma_R \gg \gamma_c$, which is the case we consider in the main text, the reservoir dynamics can be adiabatically eliminated. This behaviour is inherited by the scaled reservoir matrix elements, which are then also bound to the condensate evolution such that the equations for $\{\dot{\phi},\dot{z},\dot{\rho}\}$ are sufficient to determine condensate dynamics. We write the reservoir matrix elements $N_{nm}$ in terms of a $\phi$-dependent part $N_{nm}(\phi,z)$ and a $\phi$-independent part $N_{nm}(z)$, such that:
\begin{align}
N_{nm} = N_{nm}(\phi,z) + N_{nm}(z)
\end{align}
Then, for the diagonal reservoir matrix elements, we find:
\begin{align}
&N_{rr}(z) = \Bigg[ \frac{1-z}{2\gamma_c}\left(2A_{rrrr}\gamma_r-B_{rrrr}\gamma_c P_{\rm fr}\right) \nonumber \\
&~~~~~~~~~~~~~~~+\frac{1+z}{2\gamma_c}\left(2A_{rrll}\gamma_l -B_{rrll}\gamma_c P_{\rm fr}\right)  \Bigg]  \nonumber \\
&N_{rr}(\phi,z) = \frac{\sqrt{1-z^2}}{2\gamma_c} \Bigg[ -(\omega_l-\omega_r)(A_{rrrl}+A_{rrlr}) \sin\phi  \nonumber \\
&+\Big[ (\gamma_l+\gamma_r)(A_{rrrl}+A_{rrlr}) -\gamma_cP_{\rm fr}(B_{rrrl}+B_{rrlr} ) \Big]\cos\phi \Bigg] 
\end{align}
and:
\begin{align}
&N_{ll}(z) = \Bigg[ \frac{1+z}{2\gamma_c}\left(2A_{llll}\gamma_l-B_{llll}\gamma_c P_{\rm fr}\right) \nonumber \\
&~~~~~~~~~~~~~~+ \frac{1-z}{2\gamma_c}\left(2A_{llrr}\gamma_r - B_{llrr}\gamma_c P_{\rm fr}\right) \Bigg]  \nonumber \\
&N_{ll}(\phi,z) = \frac{\sqrt{1-z^2}}{2\gamma_c} \Bigg[ -(\omega_l-\omega_r)(A_{llrl}+A_{lllr}) \sin\phi \nonumber \\
&+\Big[ (\gamma_l+\gamma_r)(A_{llrl}+A_{lllr}) -\gamma_cP_{\rm fr}(B_{llrl}+B_{lllr} ) \Big]\cos \phi \Bigg]
\end{align}
Finally, the `off-diagonal' or `coupling' reservoir matrix element $N_{rl} = N_{lr}$ can be similarly expressed by defining:
\begin{align}
&N_{rl}(z) = \Bigg[ \frac{1-z}{2\gamma_c}\left(2A_{rlrr}\gamma_r-B_{rlrr}\gamma_c P_{\rm fr}\right) \nonumber \\
&~~~~~~~~~~~~~~+\frac{1+z}{2\gamma_c}\left(2A_{rllr}\gamma_l -B_{rlll}\gamma_c P_{\rm fr}\right) \Bigg] \nonumber \\
&N_{rl}(\phi,z) = \frac{\sqrt{1-z^2}}{2\gamma_c} \Bigg[ -(\omega_l-\omega_r)(A_{rlrl}+A_{rllr}) \sin\phi  \nonumber \\
&+ \Big[ (\gamma_l+\gamma_r)(A_{rlrl}+A_{rllr}) -\gamma_cP_{\rm fr}(B_{rlrl}+B_{rllr} ) \Big] \cos\phi \Bigg]
\end{align}
In all the above expressions, we have defined $P = P_{\rm fr}P_0$, where $P_0 = \gamma_c\gamma_R/R$. Note that this `off-diagonal' matrix element is proportional to the modal coupling at lowest order, while the `diagonal' matrix elements do in fact contain a contribution that is independent of the modal coupling.

\section{Jacobian matrix and stability}
\label{app:jac}

In this section we show that the Jacobian matrix corresponding to the system of equations for $\{\dot{\phi},\dot{z},\dot{\rho}\}$ [c.f. Eqs.~(\ref{phiFull})-(\ref{rhoFull})] takes on a particularly simple form in the regime of consideration. The equations are reproduced below, with the approximation $\gamma_l(P) \approx \gamma_r(P)$:
\begin{subequations}
\begin{align}
\dot{\phi} &= 0 =~\Delta\omega_0 -\rho G_{\phi}(\phi,z) \label{fullSysAppA} \\
\dot{z} &= 0 =~ \rho G_z(\phi,z)  \label{fullSysAppB} \\
\dot{\rho} &= 0 =~\rho\left[ (\gamma_l + \gamma_r) - G_{\rho}(\phi,z) \rho  \right] \label{fullSysAppC}
\end{align}
\end{subequations}
The associated Jacobian matrix for arbitrary $(\phi,z,\rho)$ is given by:
\begin{align}
\mathbf{J}(\phi,z,\rho) = 
\begin{pmatrix}
-\rho\partial_{\phi} G_{\phi} &  - \rho \partial_z G_{\phi} & -G_{\phi} \\
\rho\partial_{\phi} G_z & \rho \partial_z G_z & G_z \\ 
-\rho^2 \partial_{\phi}G_{\rho} & -\rho^2 \partial_z G_{\rho} & (\gamma_l+\gamma_r)-2\rho G_{\rho}
\end{pmatrix}
\end{align}
At \textit{any} fixed point of the entire system of Eqs.~(\ref{fullSysAppA})-(\ref{fullSysAppC}), the following constraints hold:
\begin{align}
G_{\phi}= \frac{\Delta\omega_0}{\rho}~,~G_z=0~,~G_{\rho} = \frac{\gamma_l+\gamma_r}{\rho} \implies (\gamma_l+\gamma_r) = \rho G_{\rho}
\end{align}
With these simplifications, the Jacobian matrix simplifies to Eq.~(\ref{jac}) of the main text, 
\begin{align}
\mathbf{J}(\phi,z,\rho)\Big|_{\rm FP} = 
\rho
\left.
\begin{pmatrix}
-\partial_{\phi}G_{\phi} & - \partial_z G_{\phi} & -\frac{\Delta\omega_0}{\rho^2} \\
 \partial_{\phi} G_z & \partial_{z}G_z & 0 \\
-\rho \partial_{\phi}G_{\rho} & -\rho \partial_z G_{\rho} & - G_{\rho}
\end{pmatrix} 
\right|_{\rm FP}
\end{align}
If we neglect the term $-\Delta\omega_0/\rho^2$, the characteristic equation $\chi=0$ for this Jacobian matrix can be easily computed:
\begin{align}
&\det \mathbf{J} = 0 = \rho^3\left(-G_{\rho}-\frac{\lambda}{\rho}\right)
\begin{bmatrix}
-\partial_{\phi}G_{\phi}-\frac{\lambda}{\rho} & - \partial_z G_{\phi} \\
 \partial_{\phi} G_z & \partial_{z}G_z-\frac{\lambda}{\rho}  \\
\end{bmatrix} \nonumber \\
&= -\lambda^3 - \lambda^2 \rho \left( G_{\rho} + \partial_{\phi}G_{\phi} - \partial_z G_z \right) +\rho^3 G_{\rho} (\partial_{\phi}G_{\phi} )(\partial_z G_z) \nonumber \\
&-\!\lambda \rho^2 \Big[G_{\rho}\partial_{\phi}G_{\phi}\!-\!G_{\rho}\partial_z G_z\! +\! (\partial_{\phi}G_z )(\partial_z G_{\phi}) \!-\! (\partial_{\phi}G_{\phi} )(\partial_z G_z) \Big] \nonumber \\
&= -\!\left(\!\frac{\lambda}{\rho}\!\right)^3\!\!\!\! -\!\left(\!\frac{\lambda}{\rho}\!\right)^2\!\!\!\! \left( G_{\rho} + \partial_{\phi}G_{\phi} - \partial_z G_z \right) + G_{\rho} (\partial_{\phi}G_{\phi} )(\partial_z G_z) \nonumber \\
&-\!\left(\!\frac{\lambda}{\rho}\!\right)\!\! \Big[G_{\rho}\partial_{\phi}G_{\phi}\!-\!G_{\rho}\partial_z G_z\! +\! (\partial_{\phi}G_z )(\partial_z G_{\phi})\! -\! (\partial_{\phi}G_{\phi} )(\partial_z G_z)  \Big] \nonumber \\
&\implies \det \mathbf{J} =0 \equiv \chi(\phi,z,\lambda/\rho)
\end{align}
In the above, we require $\rho \neq 0$, which is the physically relevant case. The final result is referenced in the main text: the characteristic equation depends on $\rho$ only via a scaling of the eigenvalues $\lambda$. As such, an increase in pump power does not change the sign of the eigenvalues, which is critical to determining stability.

\begin{figure*}[t]
\includegraphics[scale=0.4]{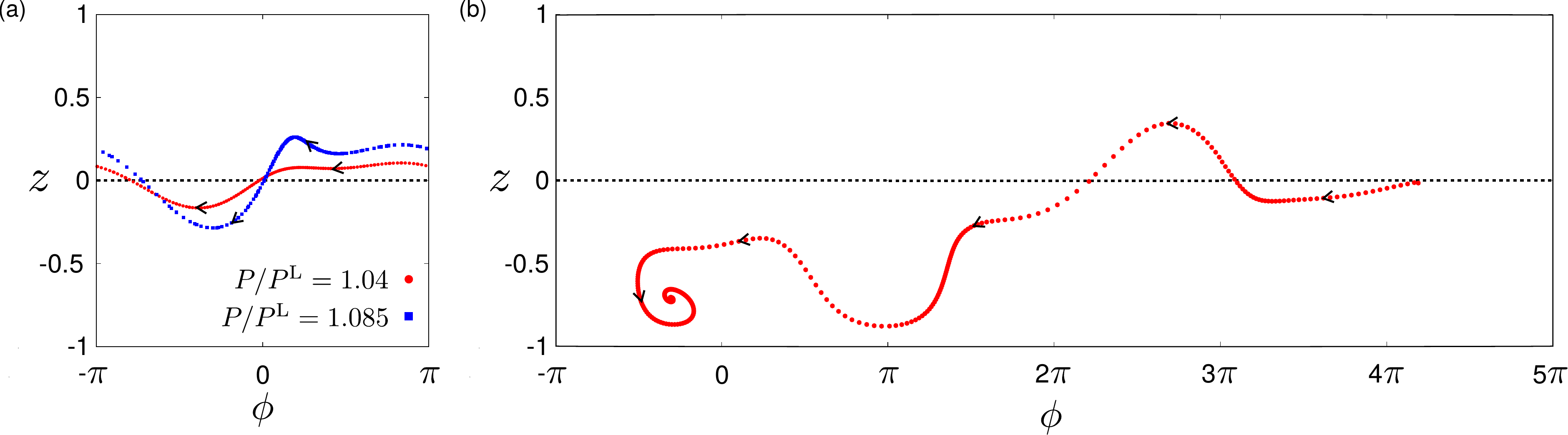}
\caption{(a) System dynamics in the $\phi$-$z$ plane in the desynchronized regime. Horizontal dashed line indicates $z = 0$. Arrows indicate the direction of flow with time; the plotted curve is repeatedly traced in the steady state. Note the large variation in $z$ over a period of the steady state oscillations. (b) System dynamics in the synchronized regime, indicating approach to the stable fixed point. Phase is unraveled to show its evolution more clearly.}
\label{sysDyn}
\end{figure*}

\begin{figure*}[t]
\begin{center}
\includegraphics[scale=0.11]{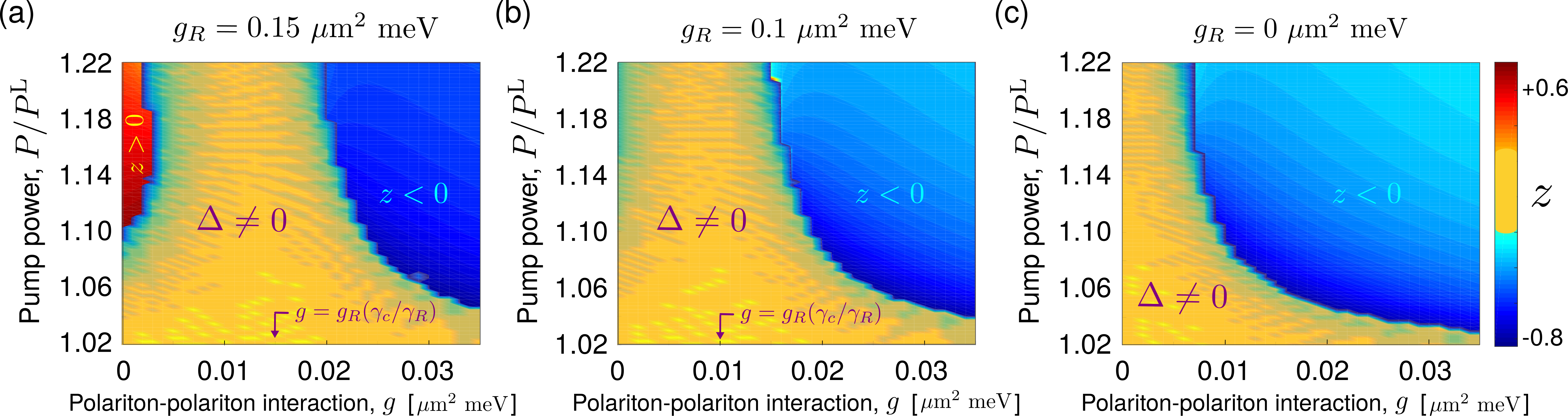}
\caption{\skc{Phase diagram in $P$-$g$ space for (a) $g_R =~0.15~\mu{\rm m^2}~{\rm meV}$, (b) $g_R =~0.1~\mu{\rm m^2}~{\rm meV}$, and (c) $g_R =~0~\mu{\rm m^2}~{\rm meV}$. The regime of $g_R$-mediated synchronization shrinks as the strength of $g_R$ decreases. Desynchronized regions are shown in yellow, while in the synchronized regions, colors indicate the steady state population imbalance $z$.}}
\label{fig:suppDet}
\end{center}
\end{figure*}

\section{Phase space system dynamics}
\label{app:sysDyn}

In this section we briefly discuss system dynamics for condensates in the coupled-trap geometry, as observed in the $\phi$-$z$ plane, in both the desynchronized and synchronized regimes. We start with initial conditions corresponding to very small initial modal occupations. The transient dynamics therefore involve a growing condensate density due to pumping, before a steady state is eventually reached. In the desynchronized regime, steady state curves are traced out in $\phi$-$z$ space, with examples shown in Fig.~\ref{sysDyn}~(a) in a typical case. Clearly, the large variation in $z$ over a period indicates the complicated amplitude dynamics can not be assumed to be approximately static. Furthermore, with increasing pump power, larger amplitude oscillations for $z$ are observed.

In the regime where a synchronized phase is possible, a stable fixed point exists as determined by the stability analysis described in the main text. Here, the system dynamically flows towards this fixed point in $\phi$-$z$ space. In the long time limit, the system localizes at this fixed point; this is shown for a typical case in Fig.~\ref{sysDyn}~(b). We show results here by unwrapping the phase $\phi$, for clarity. The exact details of the flow at a fixed pump power will depend on initial conditions, as expected.

\begin{figure*}
\begin{center}
\includegraphics[scale=0.26]{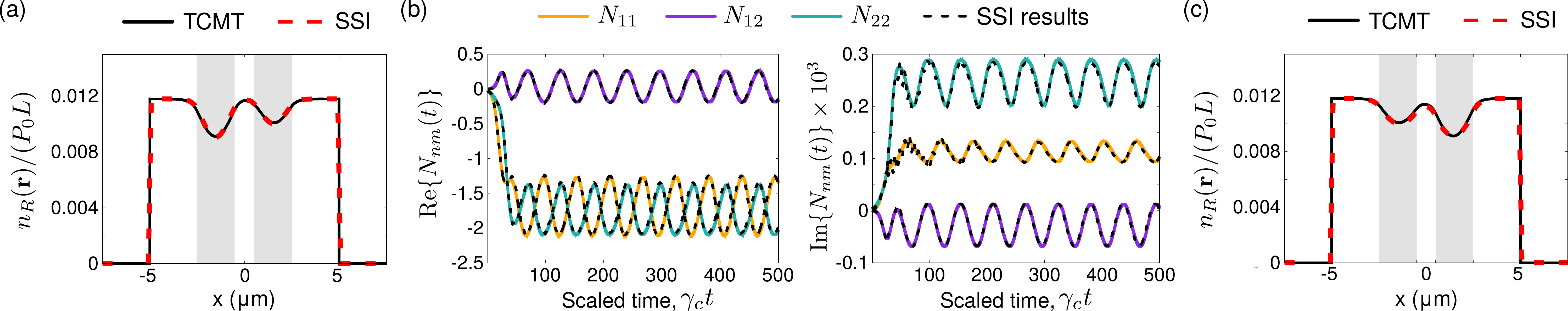}
\caption{(a), (c) Steady state reservoir density $n_R(\mathbf{r})$ scaled by $P_0 L$ (solid black) at correspondingly labelled positions in the two synchronized regions of the phase diagram in Fig.~\ref{phaseDiagComb} of the main text, together with the SSI results (dashed red). Shaded gray regions indicate the polariton traps. (b) Real and imaginary parts of reservoir matrix elements $N_{nm}(t)$ in the correspondingly labelled desynchronized region of the phase diagram, computed using the TCMT (solid lines) and SSI (black dashed lines).}
\label{fig:resDyn}
\end{center}
\end{figure*}

\skc{

\section{Supplementary simulations and reservoir dynamics}
\label{app:resDyn}

In this section, we include additional simulation results to supplement those included in the main text. We begin by presenting phase diagrams in $P$-$g$ space for values of $g_R$ weaker than the value used for Fig.~\ref{phaseDiagComb} of the main text. In Fig.~\ref{fig:suppDet}~(a), (b), and (c), phase diagrams are plotted for $g_R = (0.15, 0.1, 0)~\mu{\rm m}^2~{\rm meV}$ respectively. Clearly, the $g_R$ mediated synchronized regime shrinks as $g_R$ becomes weaker. Note, however, that the reservoir-polariton interaction $g_R$ still competes with synchronization mediated by the polariton-polariton interaction strength $g$; when $g_R$ is turned off, the $g$-mediated synchronization region is larger, over the same range of values of $g$, than the situation when $g_R \neq 0$.

In the main text we presented condensate dynamics; here, we briefly discuss the associated reservoir dynamics. In the synchronized regimes, corresponding to positions (a) and (c) in Fig.~\ref{phaseDiagComb}, the condensate has a single frequency. Both the condensate and reservoir densities become stationary in the long time limit; the latter is then obtained by solving for the steady state of Eq.~(\ref{res}):
\begin{align}
n_R(\mathbf{r}) = \frac{ Pf(\mathbf{r}) }{\gamma_R + R|\Psi(\mathbf{r})|^2 }
\label{ssRes}
\end{align}
Recall that in our modal description of condensate dynamics, the TCMT can be used to obtain $|\Psi|^2$ but generally not the full reservoir density; instead, the simulated reservoir variables are the matrix elements $N_{nm}(t)$ [c.f. Eq.~(\ref{Nnm})]. However, in the single frequency case, $n_R(\mathbf{r})$ is entirely determined by the condensate density, as is clear from Eq.~(\ref{ssRes}). Thus the TCMT can be used to obtain $n_R(\mathbf{r})$ directly in this case. We obtain the reservoir density, scaled by $P_0 L$, where $L$ is the extent of the pump, and $P_0 = \gamma_c \gamma_R /R$ as introduced earlier, for positions (a) and (c) of Fig.~\ref{phaseDiagComb}. The results are shown in solid black in Fig.~\ref{fig:resDyn}~(a) and (c) respectively, and corresponding SSI results are shown in dashed red; note the excellent agreement. For (a), the synchronized phase has $z > 0$, and so the higher frequency, left trap mode has higher occupation. This indicates the condensate density is higher in the left trap, and thus the reservoir must be more strongly depleted there. This is precisely what is observed. For (c), where $z< 0$ instead, the reservoir overlapping with the right trap is more depleted, which agrees well with the simulated results.

In the desynchronized regimes, the reservoir density has a time-dependent evolution. In this more general case, a comparison of reservoir dynamics simulated by the TCMT and SSI can be carried out via the reservoir matrix elements instead. Using the full reservoir density computed via the SSI, $N_{nm}(t)$ may be obtained using the basis modes and Eq.~(\ref{Nnm}). Comparisons of the real and imaginary parts of the matrix elements are shown in Fig.~\ref{fig:resDyn}~(b) for the TCMT (solid lines) and SSI (dashed black). Again, excellent agreement is found. Finally, we mention that for $\gamma_R \gg \gamma_c$, the reservoir evolution still adiabatically follows the condensate dynamics, and in this regime the TCMT can be used to compute the full reservoir density as well. To provide a comparison valid for more general cases, we have analyzed the reservoir matrix elements here instead.

}

\section{\skc{Weak $\gamma_R$ regime}}
\label{app:dynRes}

The dynamics we consider in this paper, and the simplified stability analysis, hold for the case where $\gamma_R \gg \gamma_c$, namely the regime where the reservoir dynamics may be adiabatically eliminated. In the opposite, dynamical reservoir regime, where $\gamma_R \ll \gamma_c$, we have found that complex dynamical effects may arise, as found in recent numerical studies~\cite{bobrovska_stability_2014, khan_non-hermitian_2016}. In particular, strongly multimode behaviour emerges for high pumping powers and - more importantly for the physics considered here - for strong nonlinearities.

\begin{figure}
\begin{center}
\includegraphics[scale=0.29]{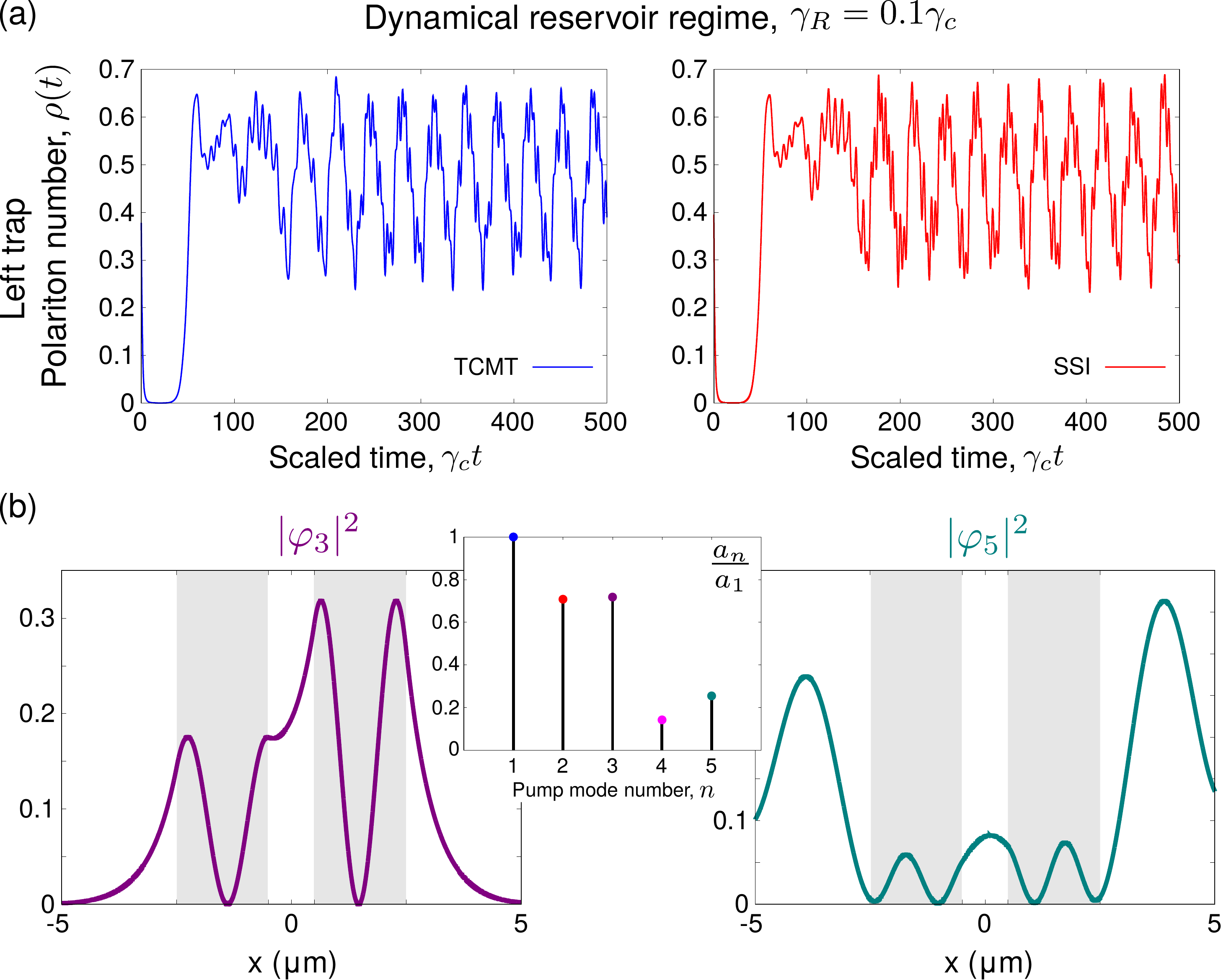}
\caption{(a) Plots of polariton number (integrated condensate density) in the left trap in the dynamical reservoir regime, $\gamma_R = 0.1\gamma_c$, using the TCMT (left panel, red) and the SSI (right panel, blue). Here, $P = 1.25P_1^{\rm L}$, $g_R = 0.2~\mu{\rm m}^2~{\rm meV}$ and $g=0.035~\mu{\rm m}^2~{\rm meV}$. (b) Inset: Spectral decomposition of condensate wavefunction, indicating the modes contributing to the solutions in (a): modes 1 and 2 two are left and right trap modes $\varphi_l$ and $\varphi_r$ respectively from the main text. Modes 3 and 5 are plotted on the left (purple) and right (turquoise) respectively.}
\label{resDynTest}
\end{center}
\end{figure}

To illustrate further the kinds of multimode dynamics present in the dynamical reservoir regime, we present in Fig.~\ref{resDynTest}~(a) plots of the polariton number (integrated condensate density) in the left trap as a function of time using the TCMT, and an exact solution using an SSI, as in the main text, Fig.~\ref{phaseDiagComb}. Here, we take $\gamma_R = 0.1\gamma_c$; the remaining parameters are summarized in the figure caption. We find complex dynamical features that agree quite well between both methods. The emergence of these features is clarified via the modal description provided by the TCMT. In Fig.~\ref{resDynTest}~(b), the inset shows a spectral decomposition of the condensate wavefunction, plotting the normalized modal weights, $\frac{a_n}{a_1}$. Only peaks corresponding to the five modes with greatest spectral weight are shown; a total of eleven modes is needed to reach the agreement shown in Fig.~\ref{resDynTest}~(a). The first two modes are simply the left and right trap modes, $\varphi_l$ and $\varphi_r$ respectively. The two modes with next greatest spectral weight, modes number 3 and 5, are depicted in Fig.~\ref{resDynTest}~(b). Hence, in this regime where reservoir relaxation is slow relative to polariton loss, condensate dynamics can typically become very complicated, and may involve multiple interacting modes, as found here.

\begin{figure}
\begin{center}
\includegraphics[scale=0.3]{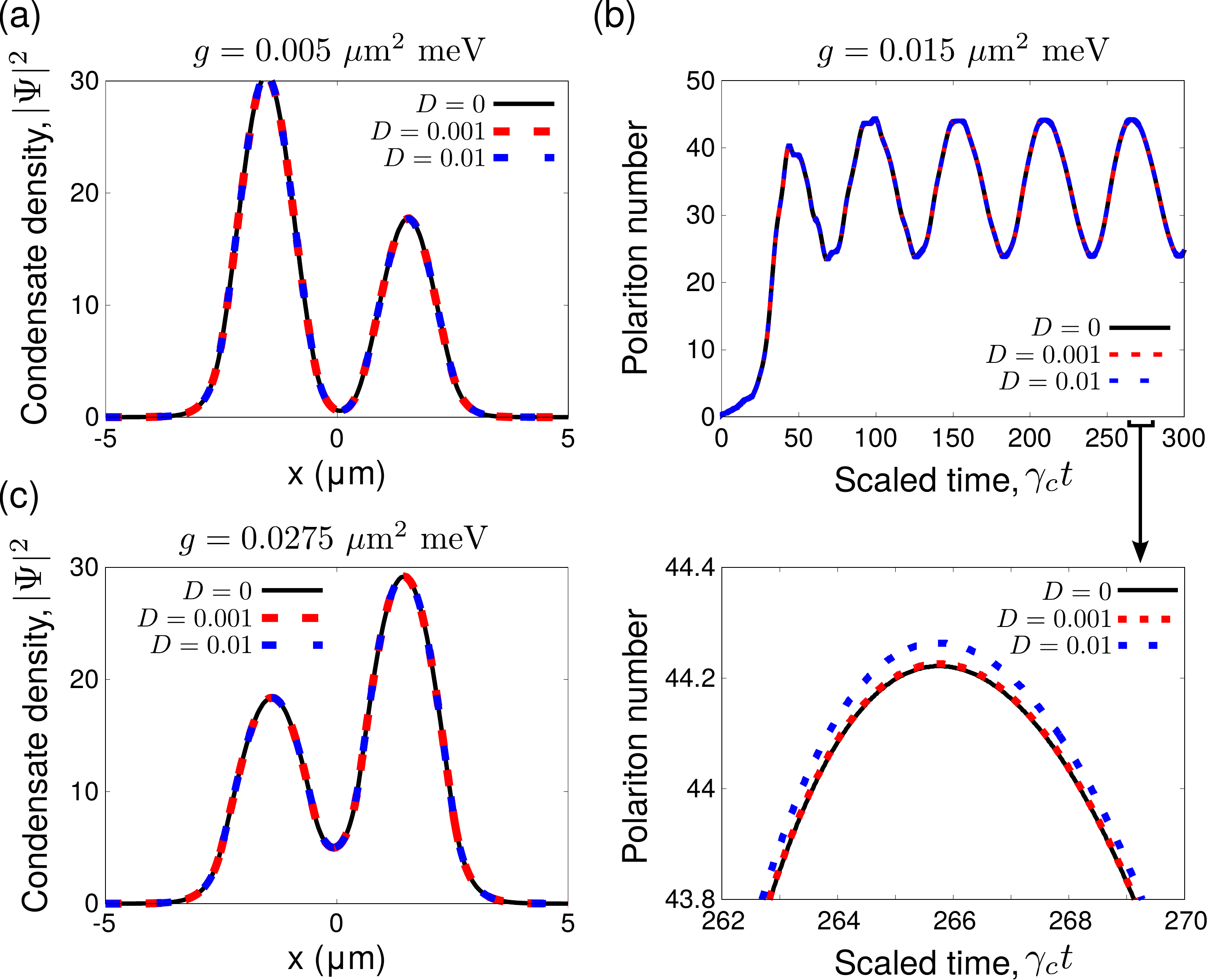}
\caption{Plots of condensate density $|\Psi|^2$ for dimensionless diffusion constants $D=0,0.001,0.01$ (black, dashed red and dashed blue respectively, in the synchronized regime, for (a) $g=0.005~\mu{\rm m}^2~{\rm meV}$ and (c) $g=0.0275~\mu{\rm m}^2~{\rm meV}$. (b) Top: Plot of the polariton number in the left trap in the desynchronized regime, $g=0.015~\mu{\rm m}^2~{\rm meV}$, for the same values of $D$ as (a) and (c). Bottom: Zoomed version of the top panel for $\gamma_c t\in [262,270]$, indicating the very minor effect of including nonzero diffusion.}
\label{diffTest}
\end{center}
\end{figure}

\section{Neglecting exciton diffusion}
\label{app:diff}

The reservoir dynamical equation in Eq.~(\ref{gGPE}) typically includes an additional term incorporating the diffusion of excitons,
\begin{align}
\partial_t n_R = Pf(\mathbf{r}) - \gamma_R n_R - Rn_R|\Psi|^2 + \frac{D}{m}\nabla^2n_R
\label{resDiff}
\end{align}
where $D$ is introduced as the \textit{dimensionless} diffusion constant. In this paper, we have neglected this term, owing to this diffusion constant being relatively small for excitons due to their heavy mass; in typical systems~\cite{wouters_excitations_2007} that we are considering here, $D \approx 10^{-3}$, which corresponds to actual diffusion rates on the order of $10~{\rm cm/s}^2$. 

However, in the present case where reservoir depletion plays an important role in determining synchronization dynamics, one may reasonably ask whether even relatively weak diffusion could strongly affect the observed physics. To verify that neglecting the diffusion term is valid, we first consider a perturbative solution in $D$; the $D=0$ reservoir dynamical equation can be solved when $\gamma_R \gg \gamma_c$, i.e. in the regime where the reservoir adiabatically follows the condensate evolution, to yield:
\begin{align}
n_R^{(0)} = \frac{Pf(\mathbf{r})}{\gamma_R + R|\Psi^{(0)}|^2}
\label{nR0}
\end{align}
The superscript $(0)$ indicates results computed with the diffusion term neglected ($D=0$). If we now expand the full reservoir density for nonzero $D$ as a power series in $D$, $n_R = n_R^{(0)} + D n_R^{(1)} + \ldots$, a dynamical equation for the first order correction $n_R^{(1)}$ to first order in $D$ is easily obtained:
\begin{align}
\partial_t n_R^{(1)} = -\gamma_R n_R^{(1)} - Rn_R^{(1)}|\Psi^{(0)}|^2 + \frac{1}{m}\nabla^2 n_R^{(0)}
\end{align}
In deriving the above, we have neglected the finite extent of the pump. Also, any modifications of $|\Psi|^2$ due to exciton diffusion will lead to contributions that are second order in $D$; hence these are neglected. Again, in the regime of adiabatic elimination, the above equation may be solved:
\begin{align}
n_R^{(1)} = \frac{1}{ \gamma_R + R|\Psi^{(0)}|^2 } \cdot \frac{1}{m}\frac{\partial^2}{\partial x^2} n_R^{(0)}
\end{align}
The effect of this correction to the diffusionless value $n_R^{(0)}$ can be estimated by computing the dimensionless ratio  $v$ defined as:
\begin{align}
v = \frac{ \int d\mathbf{r}~|\Psi^{(0)}|^2 D n_R^{(1)} }{ \int d\mathbf{r}~|\Psi^{(0)}|^2 n_R^{(0)} }
\end{align}
We compute the above quantity for a typical synchronized solution, and find $v \approx 0.01 D$. This is a very small quantity, for which we expect almost no change with the inclusion of the diffusion term.

To further confirm the results of the above analysis, we perform SSI simulations of the gGPE while retaining the diffusion term in Eq.~(\ref{resDiff}). We compute the numerical results for the three cases indicated in Fig.~\ref{phaseDiagComb} (a)-(c), and compare with the results for $D=0$. The results are computed for $D = 10^{-3}$, and an order of magnitude stronger diffusion, $D=10^{-2}$. The resulting plots are included in Fig.~\ref{diffTest} (a)-(c). We see a negligible difference between the $D=0$ and $D\neq 0$ case; the bottom panel of Fig.~\ref{diffTest} (b) shows a zoomed in version of the desynchronized dynamics in the top panel, highlighting the minute discrepancy. The $D=10^{-3}$ case is barely distinguishable from $D=0$, while the use of a stronger diffusion constant also affects the dynamics only marginally. Certainly, no qualitatively different behaviour is observed. These simulations support the omission of the diffusion term in the reservoir dynamics for the systems considered here.

\bibliography{PolSynchNew2.bib}

\end{document}